\documentclass[journal,10pt,twocolumn]{IEEEtran}
\usepackage{graphicx}
\usepackage{epsfig}
\usepackage{amssymb}
\usepackage{graphics}
\usepackage{algorithm}
\usepackage{algorithmic}
\usepackage{multirow}
\usepackage{amsmath}
\usepackage{subfigure}
\usepackage{times}
\usepackage{setspace}
\usepackage{theorem}
\usepackage{cite}
\usepackage{framed}

\ifCLASSINFOpdf

\else

\fi

\begin{document}
\title{An Electric Vehicle Charging Management Scheme Based on Publish/Subscribe Communication Framework}

\author{Yue Cao,
          Ning Wang,~\IEEEmembership{Member,~IEEE},
          George Kamel,~\IEEEmembership{Member,~IEEE} and
          Young-Jin Kim

\thanks{Y.Cao, N.Wang, G.Kamel are with the Institute for Communication Systems (ICS), University of Surrey, Guildford, UK. email: y.cao; n.wang; g.kamel@surrey.ac.uk. Y.-J. Kim is with Bell Labs, Alcatel-Lucent, Murray Hill, USA, e-mail: young.jin$\_$kim@alcatel-lucent.com.}
\thanks{The funding leading to this work is from EU FP7 C-DAX project.}
\thanks{Manuscript received 7 October, 2014; revised 20 January, 14 April, 29 May, 2015; accepted 17 June, 2015.}
}

\markboth{IEEE Systems Journal}%
{}

\maketitle

\begin{abstract}
Motivated by alleviating CO$_2$ pollution, Electric Vehicle (EV) based applications have recently received wide interests from both commercial and research communities by using electric energy instead of traditional fuel energy.
Although EVs are inherently with limited travelling distance, such limitation could be overcome by deploying public Charging Stations (CSs) to recharge EVs battery during their journeys.
In this paper we propose a novel communication framework for on-the-move EV charging scenario, based on the Publish/Subscribe (P/S)
mechanism for disseminating necessary CS information to EVs, in order for them to make optimized decisions on where to charge.

A core part of our communication framework is the utilization of Road Side Units (RSUs) to bridge the information flow from CSs to EVs, which has been regarded as a type of cost-efficient communication infrastructure.
Under this design, we introduce two complementary communication modes of signalling protocols, namely Push and Pull Modes, in order to enable the required information dissemination operation. Both analysis and simulation show the advantage of Pull Mode, in which the information is cached at RSUs to support asynchronous communication.
We further propose a remote reservation service based on the Pull Mode, such that the CS-selection decision making can utilize the knowledge of EVs' charging reservation, as published from EVs through RSUs to CSs.
Results show that both the performance at CS and EV sides are further improved based on using this anticipated information.
\end{abstract}


\IEEEpeerreviewmaketitle

\section{Introduction}
\IEEEPARstart{T}{H}E attention towards a greener environment leads to the emergence of a next generation power distribution grid, the Smart Grid.
One of the applications in Smart Grid \cite{6627967,6705637,6552997} is Electric Vehicles (EVs) \cite{doi:10.1080/00139157.2010.507143}, due primarily to their complete avoidance of CO$_2$ emissions compared to traditional fuel based vehicles.
Although EVs will represent a sizeable portion of the US national transportation fleet, with around 50$\%$ of new electric car sales by 2050, applying EVs will pose new challenges to the electricity grid, particularly in terms of EV charging management.

In contrast to previous works which investigate charging scheduling for EVs parking at home, our research interest targets to manage the charging for on-the-move EVs, relying on public Charging Stations (CSs) to provide charging services during journeys.
These public CSs are typically deployed at places where there is high concentration of EVs such as shopping mall and parking places.
In this case, on-the-move EVs requiring charging services will travel towards appropriate CSs for charging, considering as a problem on where to charge.

Different from privacy issue in traditional Smart Grid concerning Privacy-Preserving Power (PPP) request scheme to fulfill the security requirements\cite{6636035}, under EV charging scenario the concern is to not disclose the EV's private information, such as location \cite{6566195}.
In previous works \cite{Yang20132873,6651804}, on-the-move EVs normally send charging request to global controller, such that the controller will make decision on where to charge.
During this procedure, the status of EV such as location and ID will be inevitably released.
The privacy for on-the-move EV charging scenario is essential, as malicious business may bombard an individual EV with unsolicited product or service in relation to EV's location.
With this in mind, we propose a flexible charging management scheme, in which each on-the-move EV will locally make its individual decision rather than relying on the decision returned from the global controller.
In light of this, the advantage of our proposed scheme is that the status information of EV will not be released through any communication, since the decision is only made at EV side.

In order to achieve optimized charging performance such as minimizing waiting time at the EV side (for how long an incoming EV needs to wait for charging) and also balanced load across multiple CSs, the necessary information about the CS conditions can be disseminated to on-the-move EVs as an input for making their charging decisions.
Considering a given CS-selection scheme such as to minimize waiting time \cite{Yang20132873} at EV side, the freshness of CS status information received by EVs plays an important role on  charging performance.
For example, if the received information about the estimated waiting time at each CS is substantially outdated, then those EVs using such obsolete information might make inappropriate decisions.

The proposed communication framework is based on the Publish/Subscribe (P/S) \cite{Eugster:2003:MFP:857076.857078} mechanism, as a suitable communication paradigm for building applications in Vehicular Ad hoc NETworks (VANETs) with a highly dynamic and flexible nature.
Considering the on-the-move EV charging application, the P/S is also applicable where each CS as a publisher publishes its own status  information including queuing time, location, supply price, and capabilities (i.e., charging speed per unit-energy), to EVs as subscribers of the information.
Along with this, strategically deployed Road Side Units (RSUs) can support information dissemination as used by EV charging operations \cite{6649392}, through the Vehicle-to-Infrastructure (V2I) communication.
Considering that V2I communication has been mature in existing VANETs, it is worth noting that future Intelligent Transportation Systems (ITS) will necessitate wireless V2I communication for EV charging perspective in addition to road safety perspective.
In particular, how to realize RSU functions has been discussed in many previous works \cite{6459649}.

With the above in mind, we present an efficient P/S communication framework for disseminating the status information of CSs.
Here, two communication modes are introduced, namely Push Mode and Pull Mode.
In our previous work \cite{iccve}, the analysis and evaluation about the these two modes have been previously presented, focusing on the information publication frequency that will affect charging performance.
This piece of work introduces the a pioneer approach for comprehensively managing EV charging service during their journeys, including both the protocol design and the analysis on RSU based communication resource provisioning concerning EV mobility. The scenario considered here is deemed to be more challenging compared to most of the existing works \cite{6919255} that target charging management while EVs are in parking mode.

Here, we have made the following new contributions in this article:

\textbf{Communication Infrastructure Provisioning:} If the radio coverage is not ubiquitous (e.g., the WiFi scenario in which the communication while on-the-move is disruptive), an EV may miss the published information while it traverses the radio coverage of that RSU (depending on the CS publication frequency), thus affecting the received information freshness.
As such, the deployment of RSU is also linked to the control of CS information publication frequency for optimizing the charging performances.
We evaluate the influence of radio coverage and RSU numbers on charging performance, followed by the performance regarding varied EVs' speed (in relation to arrival rate) and CSs' charging power (in relation to service rate).
We then discuss how RSU resource provisioning and CS condition publication control affect the actual charging performance.

\textbf{Remote Reservation for Smart Charging Management:} Considering that the caching nature under the Pull Mode is beneficial for improving information freshness at EV side through asynchronous communication, we propose a remote reservation service based on this mode.
Here, those EVs which are in the status of travelling towards their selected CSs for charging, will publish their reservation information to these given CSs.
Such reservation information publication from EVs to CSs are also bridged by RSUs.
In particular, each CS will publish its predicted condition information according to the received EV reservations, and such information publication have anonymous EV IDs and their current locations considering the privacy issue.
This anticipated information including when an EV will arrive and how long it will need to fully recharging its battery, are used for other on-the-move EVs to estimate the expected waiting time at a CS in the near future.

\textbf{ETSI Protocols Support:} Regarding the practicality considerations, we further discuss how the ``ETSI TS 101 556-1: Electric Vehicle Charging
Spot Notification Specification'' \cite{etsi1} and the ``ETSI TS 101 556-3: Communications System for the Planning and Reservation of EV Energy Supply Using Wireless Networks'' \cite{etsi3} support our proposals.
To the best of our knowledge, this paper is the first piece of research work based on this two ETSI standards.
Here, we specifically detail technique aspects regarding the communication efficiency according to the proposed publish/subscribe communication framework.

The rest of the article is organized as follows.
In Section \uppercase\expandafter{\romannumeral2} we present the related work, followed by Section \uppercase\expandafter{\romannumeral3} in which we introduce the overall system design.
In this section we specify the design of the two P/S based communication modes for enabling communications in order to support EV charging, namely the Push Mode and the Pull Mode.
Then we evaluate the two modes based on a common-practice decision making logic for CS-selection.
The Advanced Pull Mode with remote reservation service is presented in Section \uppercase\expandafter{\romannumeral4}, followed by conclusion made in Section \uppercase\expandafter{\romannumeral5}.

\section{Related Work}
On one hand, most of previous works focus on saving charging cost, to minimize peak loads and flatten aggregated demands.
For instance, two decentralized control strategies \cite{6161220,5717547} are proposed for EV charging that establish a charging schedule to fill the overnight demand valley.
Further works \cite{5735676, eps346858, 6149116, CitationKey} consider pricing issue, in particular the work in \cite{CitationKey} brings a spatial price component induces both a temporal and spatial shift of charging activity which mitigates the load spikes.

On the other hand, few works have addressed the problem of CS-selection to alleviate the user's discomfort, by minimizing the waiting time.
The work in \cite{Yang20132873} relies on a global control center connected to all CSs, such that EVs requiring for charging will send request to obtain the status information of CSs.
The work in \cite{6183171} compares the schemes to select CS based on the closest distance and minimum queuing time, where results show that the latter performs better given high density of EVs for charging.
In \cite{Qin:2011:CSM:2030698.2030706}, CSs are considered to relay the information such as queuing time or EV reservation for charging scheduling, where the route information has been taken into account for performance optimization.
In \cite{6651850}, the CS with a higher capability to accept charging for on-the-move EV will advertise this service with a higher frequency, while EV senses this service with a decreasing function of its current level of battery. In light of this, the EV with a less battery volume will more frequently sense the service from CS.
The CS-selection scheme in \cite{6687944} adopts a pricing strategy to minimize congestion and maximize profit, by increasing or decreasing the price with the number of EVs charging at each time point.
Another work \cite{cyber} proposes a smart grid communication architecture for energy management of EVs, where the route planning and CS-selection consider battery replacement, traffic congestion on the road.
Note that previous works on CS-selection can usually be integrated with route planning issue, such as the work in \cite{eps351022} which predicts congestion at CSs and suggests the most efficient route to users.

To the best of our knowledge, no previous works have adopted the P/S mechanism to disseminate infrastructure information for EV charging application.
In this article, we utilize the RSU to bridge the information from CSs to EVs, rather than relying on their direct communication via cellular network connection.
Contrary to classical host based communication mechanism which uses location specific IP addresses to identify a receiver, the P/S mechanism allows event distribution from publisher (event producer) to subscriber (event consumer) without the use of any explicit IP address.
Here, the event distribution is based on declared subscribers' interests.
This mechanism mainly offers communications decoupled in space that subscribers do not need
to know the IP address of publishers and vice-versa), and potentially in time if the system is able to store events for clients which are temporally disconnected, such as the intermittent connection resulting from rapid topology changes and sparse network density in Delay/Disruption Tolerant Networks (DTNs) \cite{my}.

\section{Proposal Of EV Charging Management Using P/S Communication Framework}

\subsection{Overview of Push and Pull Modes}
\begin{figure}[htbp]
  \centering
  \subfigure[Push Mode]
  {
  \centering
    \label{push}
    \includegraphics[scale=0.3]{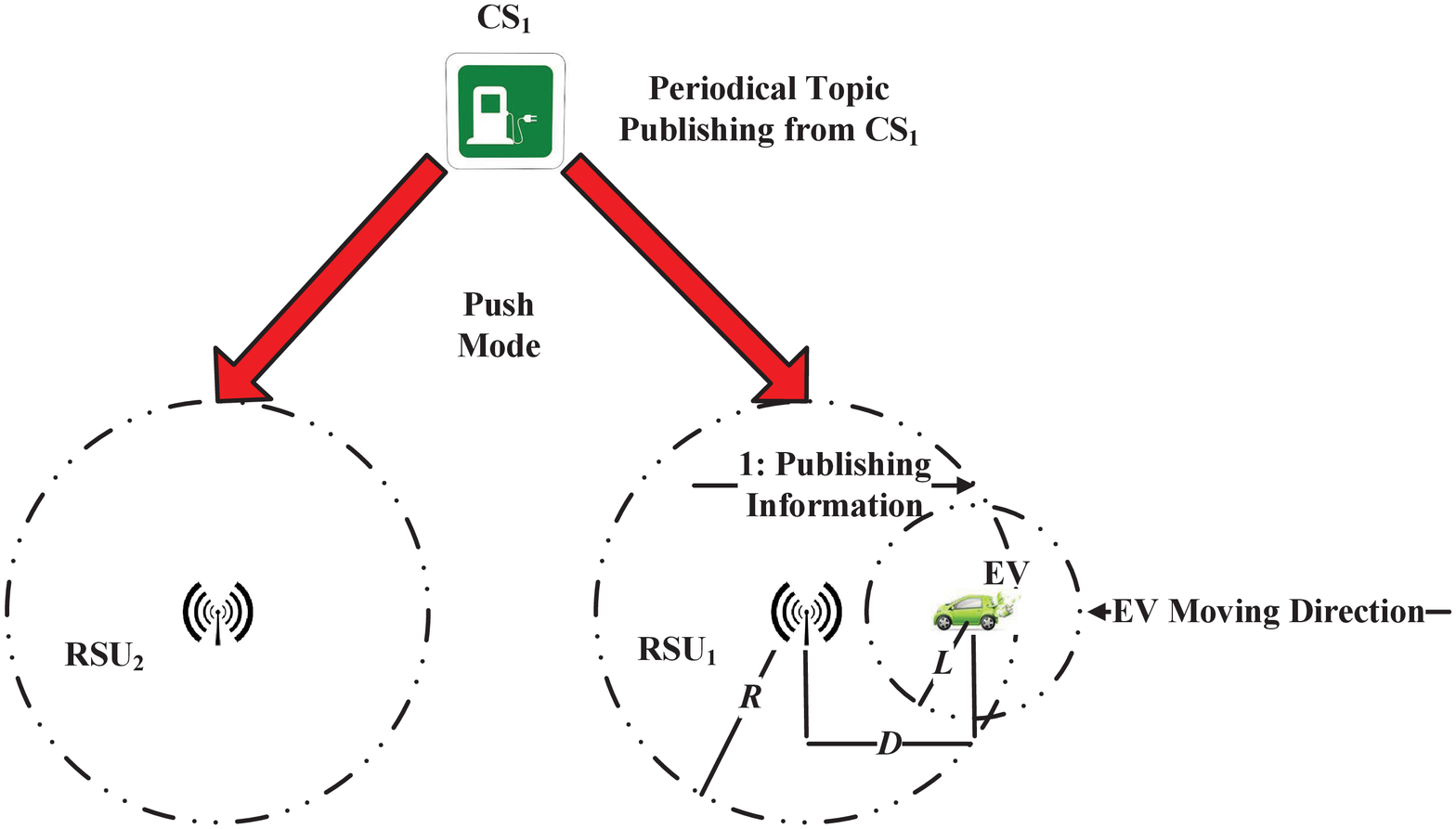}
    }
  \subfigure[Pull Mode]
  {
  \centering
    \label{pull}
    \includegraphics[scale=0.3]{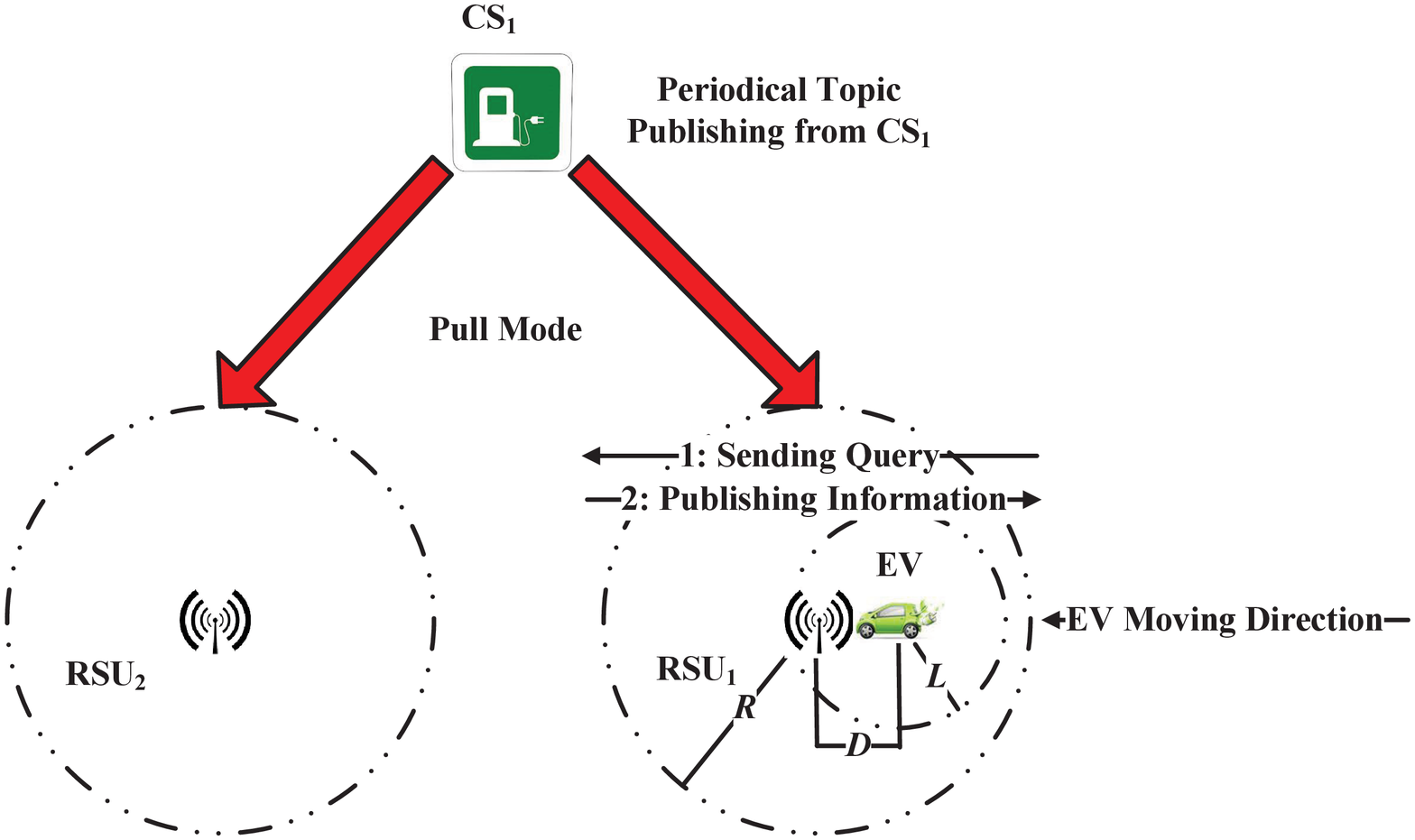}
    }
  \caption{Two Communication Modes}\vspace{-5pt}
\end{figure}

The ``ETSI TS 101 556-1'' \cite{etsi1} introduces several entities to support on-the-move EV charging scenario.
The basic application is to notify EV drivers about the CS condition information, and then the driver is able to select a CS (based on the calculated minimum CS queuing time considering other EVs locally under charging and waiting for charging at a CS) for re-charging his EV. Our proposed Push/Pull Mode provides an efficient communication manner for this information notification purpose.

\textbf{Push Mode:} Under the Push Mode as shown in Fig.\ref{push}, the EV, as subscriber, passively receives information from a nearby RSU.
This happens when the EV is within the radio coverage of that RSU, as given by $(D\leq R)$.
Here, $D$ is the distance between RSU and EV, while $R$ is the radius of RSU radio coverage.
Note that the RSU under this mode will not cache any historical information received from a CS, thus a communicating EV can not obtain any information if the CS is not currently publishing its information.

\textbf{Pull Mode:} Under the Pull Mode, each RSU locally caches the information from a CS as the historical record.
The EV which has a receiving range of $L$, initially sends an explicit query to the RSU, when their current distance is smaller than the minimum value between their radio coverage, as given by $(D\leq \min [R,L])$.
In general, we consider $(L<R)$.
Upon receiving this query, the RSU then sends its latest cached information to that EV, as shown in Fig.\ref{pull}.
Note that once a new value has been received, it will replace the obsolete values in the past, that are not necessarily maintained by RSUs.
In contrast to the Push Mode in which the information is received multiple times from each RSU, under the Pull Mode, each EV can only obtain the information from a RSU only once.

\begin{figure}[htbp]
\begin{center}
\includegraphics[width=9cm,height=3.5cm]{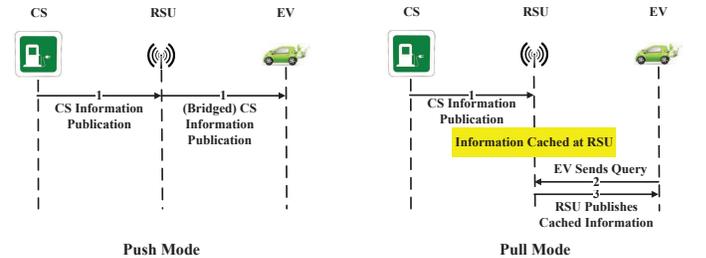}
\caption{Timing Sequences for Push and Pull Modes}\vspace{-10pt}
\label{sequence}
\end{center}
\end{figure}

\textbf{Time Sequences:}
Under both modes, we assume all CS information publication are synchronized, while RSUs could aggregate all information from multiple CSs and send them to EVs for one time only.
Such a way facilitates an efficient radio resource utilization and alleviated interference to EVs.
Therefore, the CS information is firstly published to RSUs, then the aggregated information for all CSs at RSUs side, will be obtained by those EVs passing through RSUs.
Here, either Push or Pull Mode can be applied, while their timing sequences are shown in Fig.\ref{sequence}.

\subsection{Analysis}
We assume all EVs are able to obtain the location of each CS directly via the navigation system, and the CS has sufficient electric energy for charging all the time. Although our proposal is based on this assumption herein, its application is not limited for other special cases in reality when considering the limited capacity of CSs.
Our communication framework is based on the situation that each CS periodically publishes information in relation to its instantaneous queuing time.
Each EV will leave from a CS once its energy is fully charged.
For simplicity, the CS is connected to all RSUs on the road and there is no overlap between the radio coverage of adjacent RSUs, for instance under the WiFi scenario where the radio coverage is not ubiquitous.
For the purpose of analysis, we model an event that an EV leaving from a RSU and moving to next one, based on (average) constant moving speed.

\begin{figure}[htbp]
\begin{center}
\includegraphics[scale=0.35]{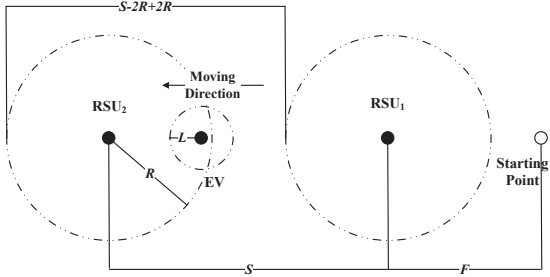}
\caption{Simplified Scenario With 2 RSUs}\vspace{-10pt}
\label{rsuanalysis}
\end{center}
\end{figure}

In Fig.\ref{rsuanalysis}, in addition to the definitions of $R$ and $L$, $T$ denotes the CS publication frequency and $V$ denotes an average constant EV moving speed.
Also, $S$ denotes the distance between adjacent RSUs, and $F$ denotes the distance between the first RSU and starting point of EV.
The detail of derivation on probability for EV to access information from at least one RSUs can be referred to \cite{iccve}.
Here, we provide final probability results depending on the two modes respectively, namely $P_{push}$ and $P_{pull}$.

Given that there are $N$ RSUs deployed, we have $P_{push}$:
\begin{equation}\footnotesize
P_{push}\leq1-\left(1-\frac{F+R}{V\cdot T}\right)\left(1-\frac{4R^2}{V\cdot T\cdot S}\right)^{(N-1)}
\end{equation}
We observe that the probability an EV receiving information depends, in general, on a larger radio coverage and larger number of RSUs.
Meanwhile, a frequent CS update interval and slower EV moving speed also improve such probability.
Recall that we consider that there is no overlap between the radio coverage of two adjacent RSUs, thus a closer distance between them is beneficial to increase $P_{push}$ as well.

Regarding $P_{pull}$, we have:
\begin{equation}\footnotesize
P_{pull}\leq1-\prod^{N}_{i=1}\left\{1-\left[\frac{(i-1)S+F+L}{V\cdot T}\right]\right\}
\end{equation}
It is observed that increasing the radio coverage of EV improves the probability to obtain information from RSUs.
Similar to that under the Push Mode, the influence of $V$ and $T$ are also applicable in this case.
However, since it is beneficial to wait for a longer time to cache the historical information of CS under the Pull Mode, a larger $S$ is desirable.

Recall that the radio coverage between adjacent RSUs is not ubiquitous, we have $(2R \leq S)$.
Then the upper bound of $P_{push}$ can be converted as:
\begin{equation}\footnotesize
\begin{split}
P^{upper}_{push}&=1-\left(1-\frac{F+R}{V\cdot T}\right)\left(1-\frac{4R^2}{V\cdot T\cdot S}\right)^{(N-1)}\\
           &\leq1-\left(1-\frac{F+R}{V\cdot T}\right)\left(1-\frac{S^2}{V\cdot T\cdot S}\right)^{(N-1)}\\
           &=1-\left(1-\frac{F+R}{V\cdot T}\right)\left(1-\frac{S}{V\cdot T}\right)^{(N-1)}
\end{split}
\end{equation}
Next, the upper bound of $P_{pull}$ can be converted as:
\begin{equation}\footnotesize
\begin{split}
P^{upper}_{pull}&=1-\prod^{N}_{i=1}\left\{1-\left[\frac{(i-1)S+F+L}{V\cdot T}\right]\right\}\\
                      &>1-\left(1-\frac{F+L}{V\cdot T}\right)\prod^{N}_{i=2}\left\{1-\left[\frac{(2-1)S+F+L}{V\cdot T}\right]\right\}\\
                      &=1-\left(1-\frac{F+L}{V\cdot T}\right)\left(1-\frac{S+F+L}{V\cdot T}\right)^{(N-1)}
\end{split}
\end{equation}
Based on the above, it is observed that if setting $(R=L)$ for fairness, the Pull Mode always achieves a higher probability to obtain information from RSUs, than Push Mode.
In light of this, the number of times that EVs obtain information under the Pull Mode is higher than that under the Push Mode, based on multiple times encounter between EVs and RSUs.

A simulation validation has been investigated to justify the above analysis, where detail can be referred to \cite{iccve}.
Note that, the straight road scenario model with constant (average) vehicle moving speed has already been maturely adopted by researches \cite{5473218} in VANETs for characterizing the analysis that is applicable under realistic scenario with a more complicated road topology and variable vehicle moving speed.
Although such analysis is based on a straight road where an EV will pass through all RSUs with fixed inter-RSU distance, the nature of the proposed Push/Pull Mode is certainly also applicable under a complex/realistic city scenario.
Here, the Push Mode mainly relies on the chance to encounter RSU meanwhile while there is currently a CS publication.
The Pull Mode relies on the event that RSUs have cached the latest published CS information in the past.
Under city scenario, EVs are with varied moving speed, and RSUs are deployed along non-straight roads with varied inter-RSU distances.
The difference of these two conditions still do not affect the nature of the proposed Push/Pull Mode, shown as the following simulation results under Helsinki city scenario.

\subsection{Decision Making Procedure for EV Charging}
All CSs are connected with each RSU through dedicated and reliable communication channels.
While each EV communicates with RSUs to gather status information about all CSs.
The on-the-move EV reaching a threshold on its residual battery charge applies a pre-defined policy to select a dedicated CS for charging, by using the information obtained from RSU.
Note that an on-the-move EV might has received information for several times when it is reaching the threshold for requesting charging.

Here, such EV selects a dedicated CS within its reachability, based on the minimum queuing time at CS.
Since EVs' decision making is always based on the latest published information, the information freshness (which affects the actual charging performance) effectively depends on how often the periodically published information is received by the on-the-move EVs.
In the worst case, the EV would select a CS with the shortest geographic distance as a back-up scheme, if none of the information in relation to any CS is obtained from RSUs.
Note that this situation typically happens when that EV misses all update when traversing RSUs' radio coverage.
Finally, upon reaching the selected CS, EVs are then scheduled based on the First Come First Serve (FCFS) priority for charging.

\begin{table}[!htb]\scriptsize
\renewcommand{\arraystretch}{1.3}
\caption{List of Notations}
\label{configure1}
\centering
\begin{tabular}{|c|c|}
\hline
$N_C$ & Number of EVs under charging at CS\\\hline
$N_W$ & Number of EVs waiting for charging at CS\\\hline
$\vartheta$ & Number of charging slots at CS\\\hline
$E^{max}_{ev}$ & Full volume of EV battery\\\hline
$E^{cur}_{ev}$ & Current volume of EV battery\\\hline
$\beta$&Charging power at CS\\\hline
$T^{arr}_{ev}$&EV's arrival time at CS\\\hline
$T^{tra}_{ev}$&EV's travelling time to reach CS\\\hline
$T^{cha}_{ev}$&Expected charging time upon arrival\\\hline
$T_{cur}$&Current time in the network\\\hline
$S_{ev}$&Moving speed of EV\\\hline
$\alpha$&Electric energy consumed per meter\\\hline
$N_R$&Number of reservation entries\\\hline
\end{tabular}
\end{table}

In order to calculate the CS instantaneous queuing time, we need the following information defined in TABLE \ref{configure1}:
\begin{itemize}
\item Number of EVs under charging, denoted by $N_C$.
\item Charging time of each EV parking at CS, as given by $\frac{E^{max}_{ev}-E^{cur}_{ev}}{\beta}$.
\item Number of charging slots at CS, denoted by $\vartheta$.
\item Number of parked EVs which are still waiting for available charging slots, denoted by $N_W$.
\end{itemize}

\begin{algorithm}[htbp]\footnotesize
\caption{Calculate the Minimum EVs' Charging Time}
\label{alg1}
\begin{algorithmic}[1]
\STATE define $\text{MINIMUM}=+\infty$
\IF {$(N_C<\vartheta)$}
\RETURN $\text{MINIMUM}=0$
\ENDIF
\FOR{$(i=1$; $i\leq N_C$; $i++)$}
\IF {$\left(\frac{E^{max}_{ev_{(i)}}-E^{cur}_{ev_{(i)}}}{\beta}<\text{MINIMUM}\right)$}
\STATE $\text{MINIMUM}=\frac{E^{max}_{ev_{(i)}}-E^{cur}_{ev_{(i)}}}{\beta}$
\ENDIF
\ENDFOR
\RETURN $\text{MINIMUM}$
\end{algorithmic}
\end{algorithm}

The calculation of instantaneous queuing time is obtained as follows:
\begin{itemize}
\item Firstly, since the number of EVs under charging can not exceed the value of $\vartheta$, the remaining time to wait for an available charging slot is equivalent to the minimum charging time of those EVs being charged, if all charging slots are occupied as presented in Algorithm \ref{alg1}.
\item Secondly, another factor from those $N_W$ number of EVs still waiting for charging is to calculate an accumulative value of their charging time, presented between lines 7 and 9 in Algorithm \ref{alg2}.
\item In the special case as presented between lines 2 and 6, one of the parked EVs will be scheduled for charging given $(N_C<\vartheta)$, indicating that there is at least one available charging slot. Therefore, charging slots will always be occupied if there are EVs parking at CS.
\end{itemize}

Based on this abstract estimation on congestion status of CS, the CS at which there is a large number of EVs parking will be estimated with a long queuing time, in particular that applying more charging slots contributes to a short queuing time due to reducing the number of $N_W$.
Note that using closed-form queuing theory assuming the EV arrival rate is known in advance and constant \cite{6651804}, will improve the charging performance. However, this application has limitation in reality where the EV arrival rate is uncertain and dynamic (as for our simulation scenario).
For each update interval, each CS will calculate its instantaneous queuing time and publish this information to on-the-move EVs through RSUs.

\begin{algorithm}[htbp]\footnotesize
\caption{Calculate CS's Instantaneous Queuing Time}
\label{alg2}
\begin{algorithmic}[1]
\STATE define $\text{VALUE}=0$
\IF {a charging slot is free}
\STATE schedule another EV (in the queue of $N_W$) for charging based on FCFS
\STATE add this EV into the queue of $N_C$
\STATE delete this EV from the queue of $N_W$
\ENDIF
\FOR{$(i=1$; $i\leq N_W$; $i++)$}
\STATE $\text{VALUE}=\text{VALUE}+\frac{E^{max}_{ev_{(i)}}-E^{cur}_{ev_{(i)}}}{\beta}$
\ENDFOR
\RETURN $\text{VALUE}+\text{Output From Algorithm \ref{alg1}}$
\end{algorithmic}
\end{algorithm}

\subsection{Performance Evaluation}
\subsubsection{Scenario Configuration}
\begin{figure}[htbp]
\begin{center}
\includegraphics[scale=0.5]{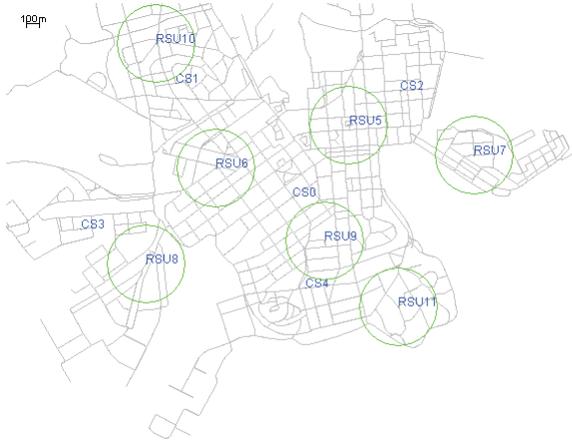}
\caption{Simulation Scenario of Helsinki City}\vspace{-10pt}
\label{hel}
\end{center}
\end{figure}

We have built up an entire system for EV charging in Opportunistic Network Environment (ONE) \cite{onesimulator}, a java based simulator  particularly developed for research on DTNs.
In Fig.\ref{hel}, the default scenario with 4500$\times$3400 $m^2$ area is shown as the down town area of Helsinki city in Finland.
Here, 100 EVs with $[30\sim50]$ $km/h$ variable moving speed are initialized in the network.
The configuration of EVs follows the charging specification (Maximum Electricity Capacity ($\text{MEC}$), Max Travelling Distance ($\text{MTD}$): 30 kWh, 161 km) of Wheego Whip EV \cite{wheego}.
Here, the electricity consumption for the Traveled Distance ($\text{TD}$) is calculated based on $\frac{\text{MEC}\times \text{TD}}{\text{MTD}}$ by referring to \cite{Qin:2011:CSM:2030698.2030706}, and we set Status Of Charge $(\text{SOC})=40\%$ for EV to start selecting CS. Here, the shortest path towards CS is formed considering road topology.

Considering these 100 EVs in network, 5 CSs are provided with 3000 kWh electric energy and 3 charging slots through entire simulation, using the fast charging rate of 62 kW.
Under this configuration, the charging management is essential as some EVs have to wait additional time for charging, until charging services for other EVs in front of the queue at a CS are finished.
For the purpose of fairness, 300m radio coverage is applied for 7 RSUs and 100 EVs, where all EVs continually receive information regardless of their SOC.
The default update interval (publication frequency) of CS is 100s while simulation time is 43200s = 12 hours.

\subsubsection{Comparable Performance Metrics}
We also evaluate the charging system based on an Ideal Case, that each EV could obtain the CS instantaneous queuing time by sending request and receiving decision reply from a global controller. This is the common framework applied by previous works, performing in a centralized way.
Our proposed Push and Pull Modes however are with a distributed nature, where the CS-selection is made by each EV locally.

We are mainly concerned with the performance affected by communication patterns, with 95$\%$ confidence interval based on 10 runs.
The evaluation metrics are as follows:
\textbf{1) Average Waiting Time - }The average period between the time an EV arrives at the selected CS and the time it finishes recharging its battery.
\textbf{2) Number of Times EVs Obtain information - }The total number of times that all EVs obtain information from RSUs.
\textbf{3) Average Information Freshness - }The average value of the difference between the current queuing time at CS side and that recorded at EV side, only calculated when an EV makes its individual selection decision.
\textbf{4) Utilization of CSs - }The amount of consumed electric energy calculated at CS side.
\textbf{5) Number of Charged EVs - }The total number of fully charged EVs in the network.

\subsubsection{Influence of Update Interval}
\begin{figure}[htbp]
  \centering
  \subfigure[Average Waiting Time]
  {
  \centering
    \label{f21}
    \includegraphics[width=4.1cm,height=2.5cm]{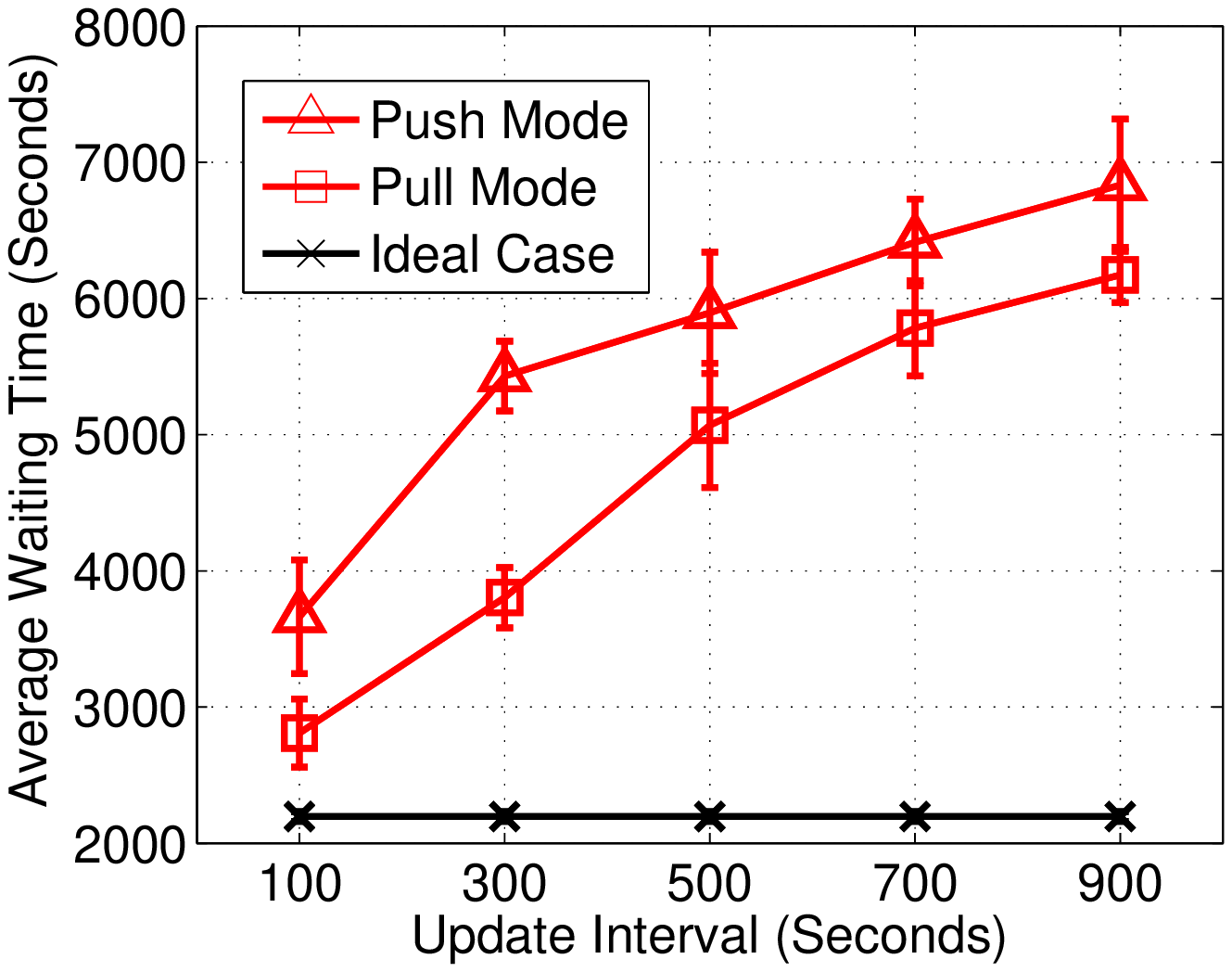}
    }
  \subfigure[Number of Times EVs Obtain Information]
  {
  \centering
    \label{f22}
    \includegraphics[width=4.1cm,height=2.5cm]{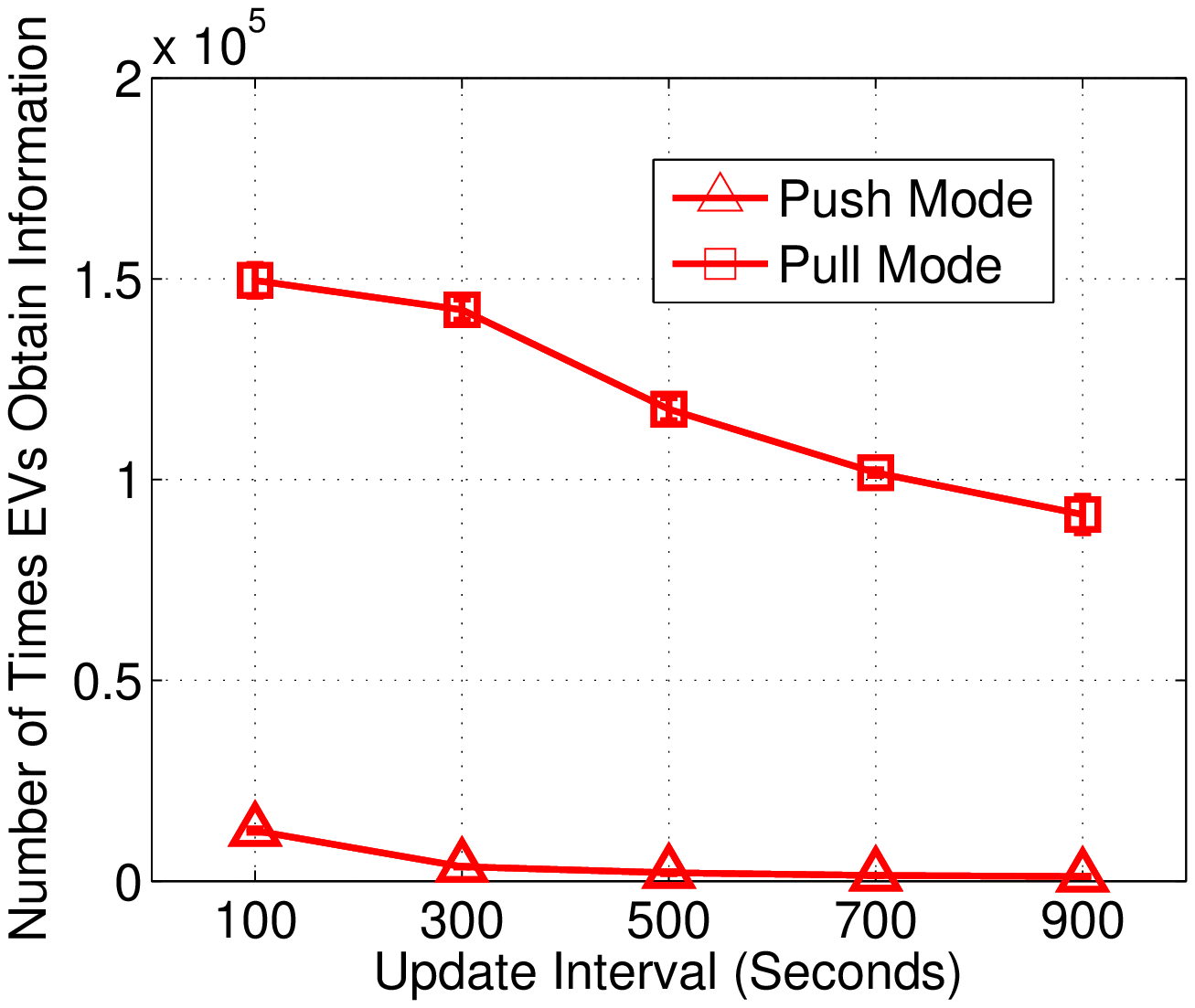}
    }
      \subfigure[Average Information Freshness]
  {
  \centering
    \label{f23}
    \includegraphics[width=4.1cm,height=2.5cm]{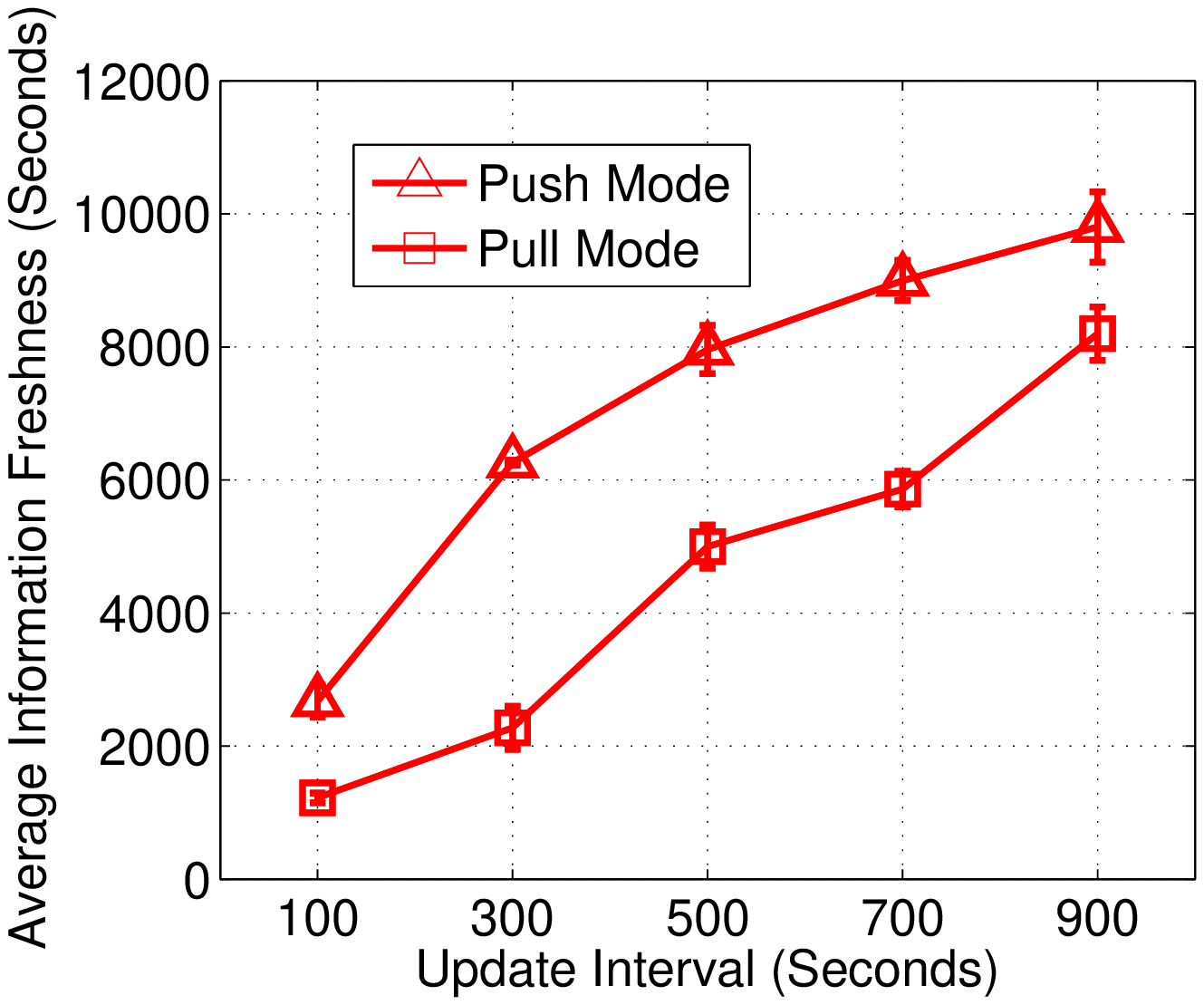}
    }
    \subfigure[Number of Charged EVs]
  {
  \centering
    \label{f24}
    \includegraphics[width=4.1cm,height=2.5cm]{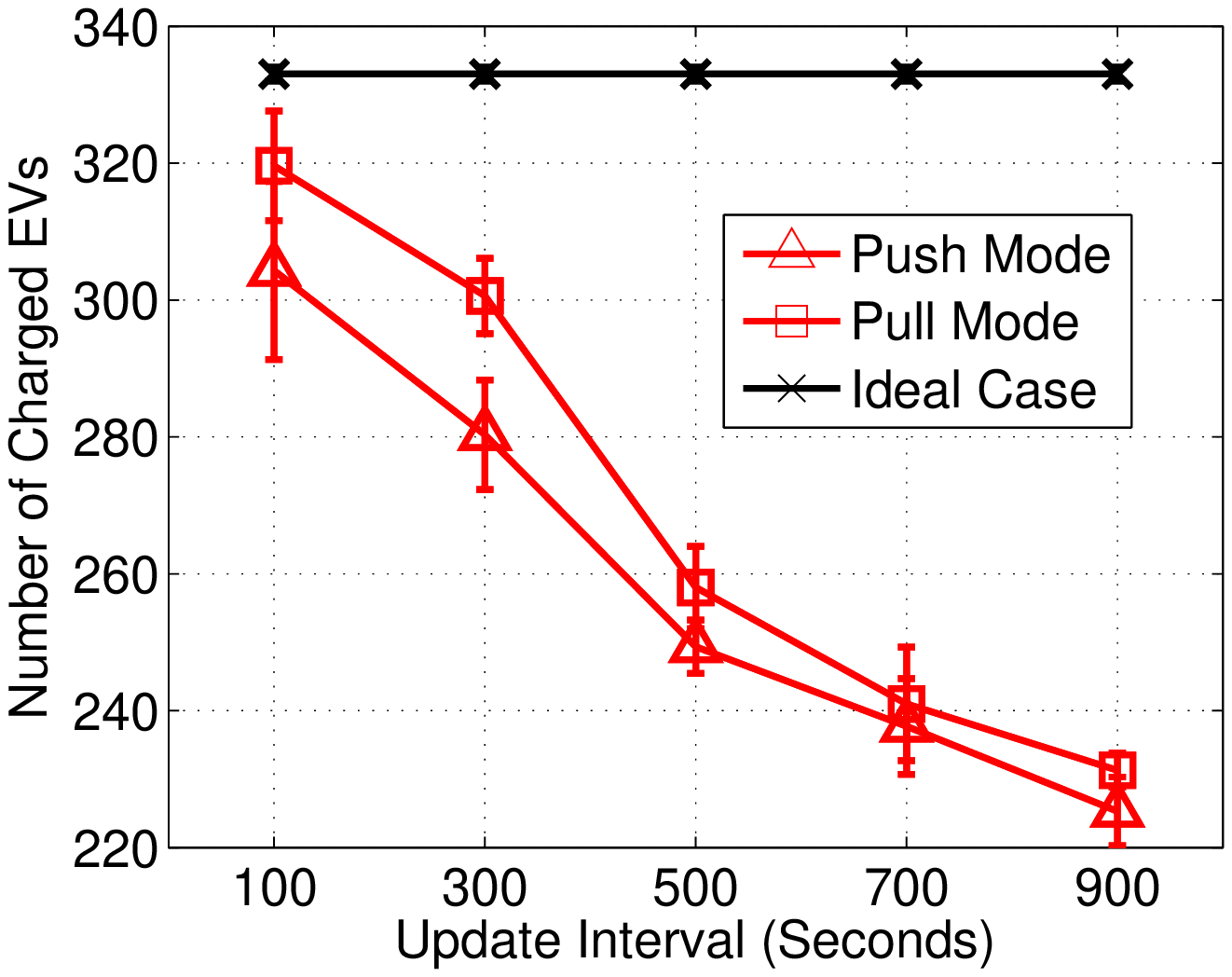}
    }
    \subfigure[Utilization of CSs]
  {
  \centering
    \label{f31}
    \includegraphics[width=6.5cm,height=2.5cm]{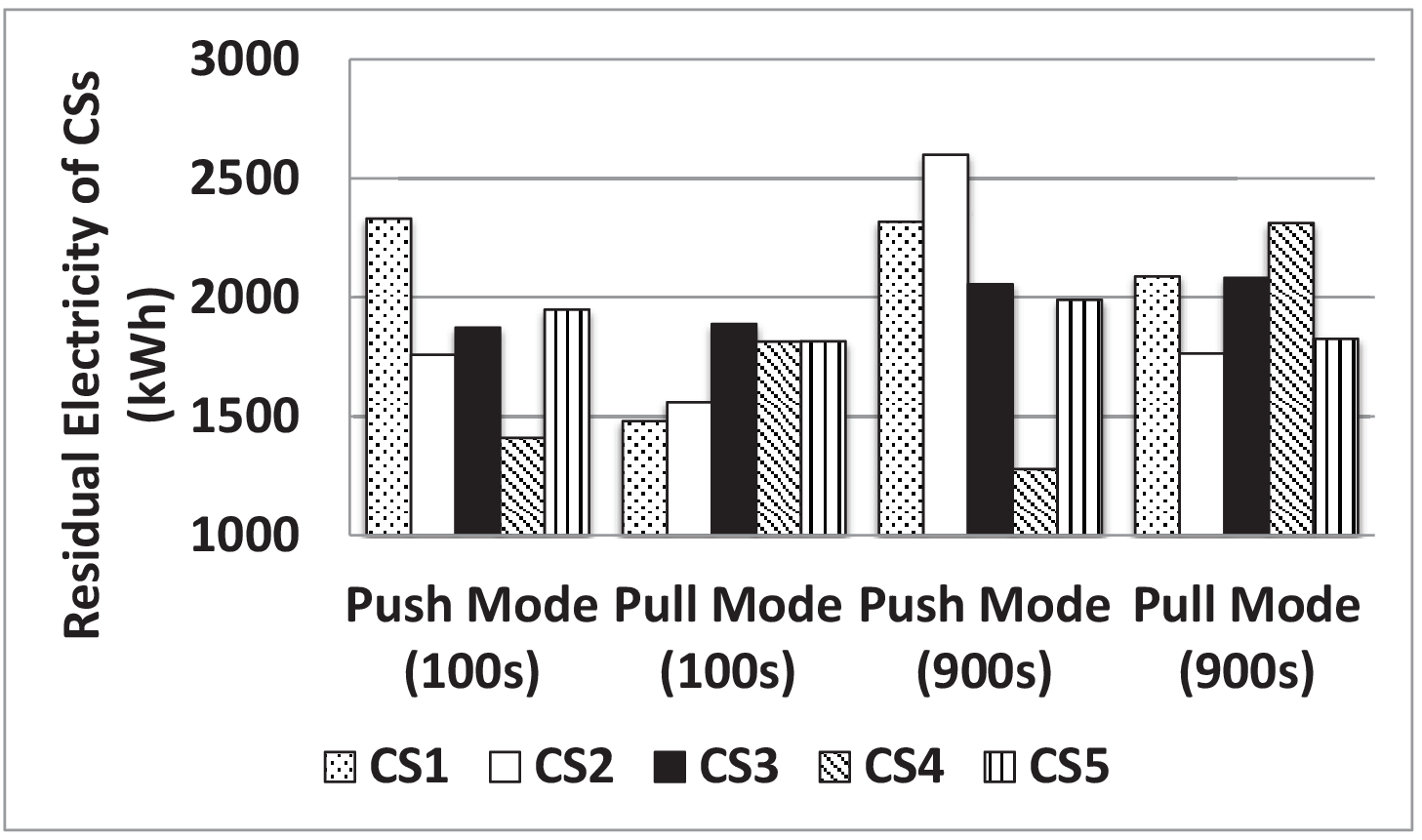}
    }
  \caption{Influence of Update Interval $T$}\vspace{-10pt}
\end{figure}

In Fig.\ref{f21}, with an infrequent update interval of CS, all EVs in the network experience an increased average waiting time.
This is due to the fact that using an outdated information affects the computation at the EV side to make CS-selection decision.
In other words, the number of EVs awaiting at CS, as estimated at the EV side when making decision, may be significantly different from that at the CS side.
Thus, with an increased update interval, there will be a huge difference between that performance given 100s and 900s intervals.
In particular, by relying on the realtime information about CSs in Ideal Case, the obtained information is the same as the status of CSs.
As such, the performance under the Ideal Case achieves the lowest average waiting time in Fig.\ref{f21}, particularly when both Push and Pull Modes are based on 900s update interval.
In Fig.\ref{f22}, the number of times EVs obtain information is decreased given a longer update interval, where this performance under the Push Mode is worse than that under the Pull Mode.
Upon this result, we observe the decreased number of times to obtain information results in poor information freshness in Fig.\ref{f23}.
Therefore, all EVs will experience a longer waiting time due to using outdated information for CS-selection, which results in a lower number of charged EVs in Fig.\ref{f24}.
In Fig.\ref{f31}, we further observe that the number of times to obtain information has influence on the utilization of CSs.
This is because that the poor information freshness yields EVs to make inaccurate selection decision, as such the electric energy at some CSs may not be utilized for charging.
The above observation becomes more significantly given 900s update interval.

\subsubsection{Influence of Radio Coverage}
\begin{figure}[htbp]
  \centering
  \subfigure[Average Waiting Time]
  {
  \centering
    \label{f41}
    \includegraphics[width=4.1cm,height=2.3cm]{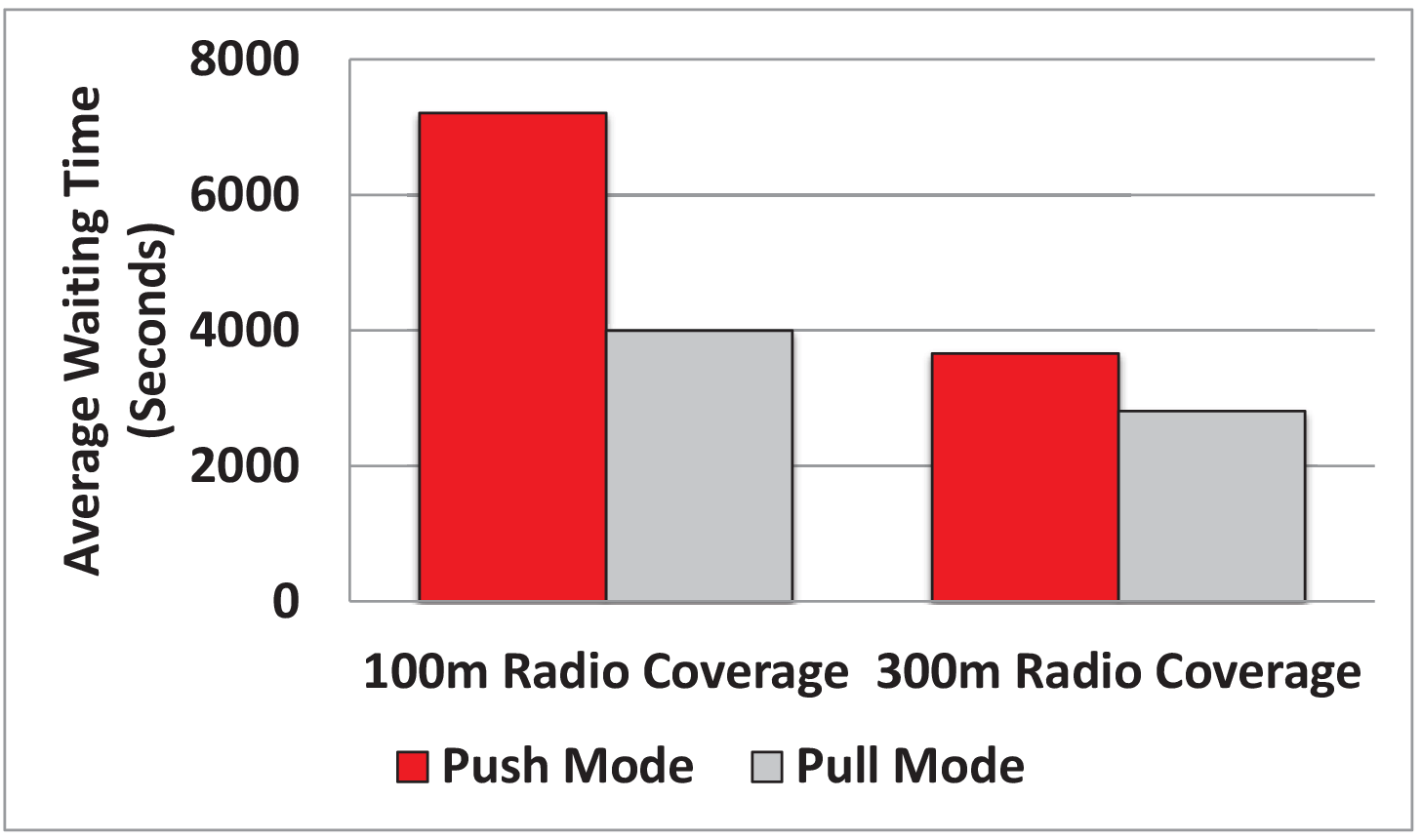}
    }
  \subfigure[Number of Times EVs Obtain Information]
  {
  \centering
    \label{f42}
    \includegraphics[width=4.1cm,height=2.3cm]{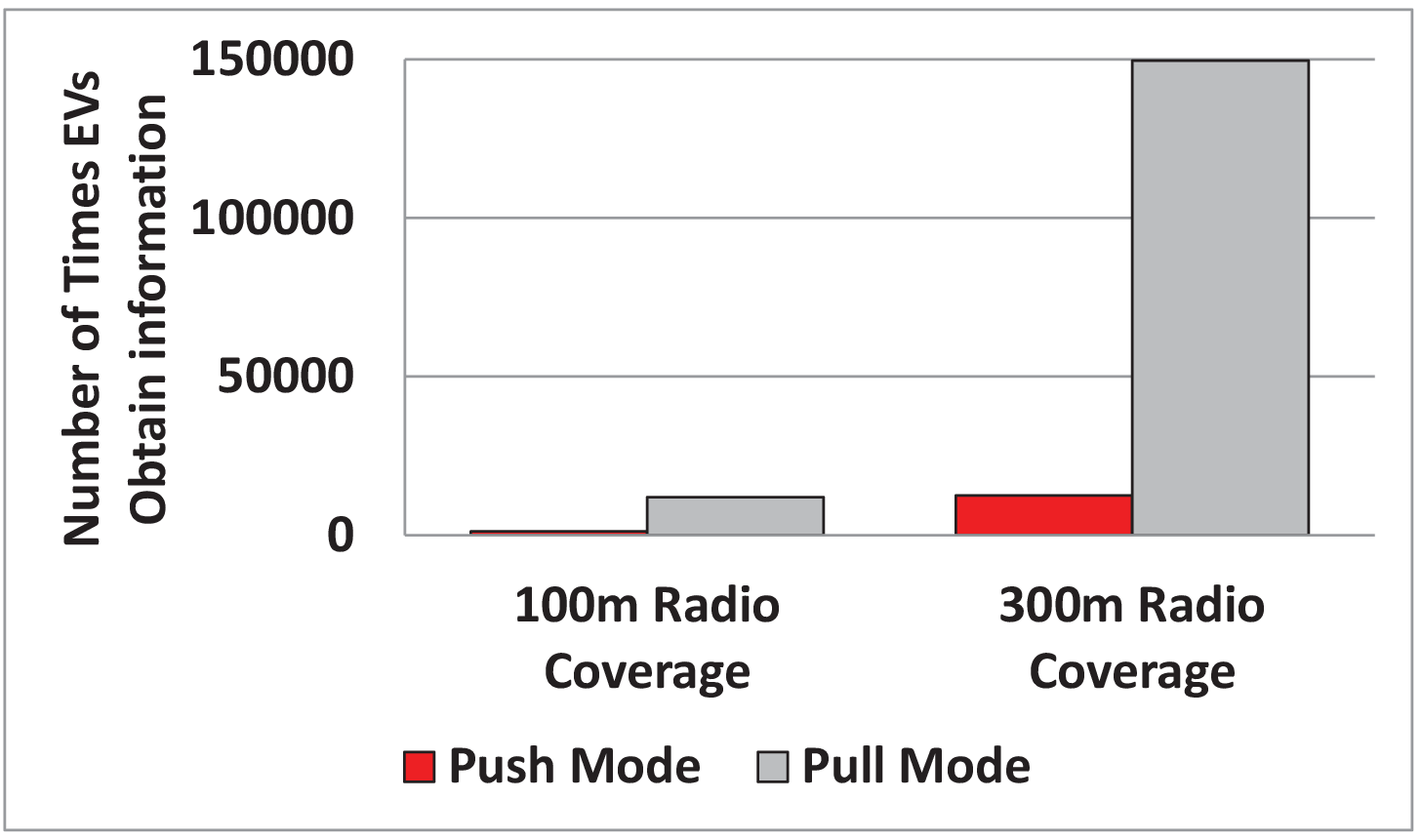}
    }
    \subfigure[Average Information Freshness]
  {
  \centering
    \label{f43}
    \includegraphics[width=4.1cm,height=2.3cm]{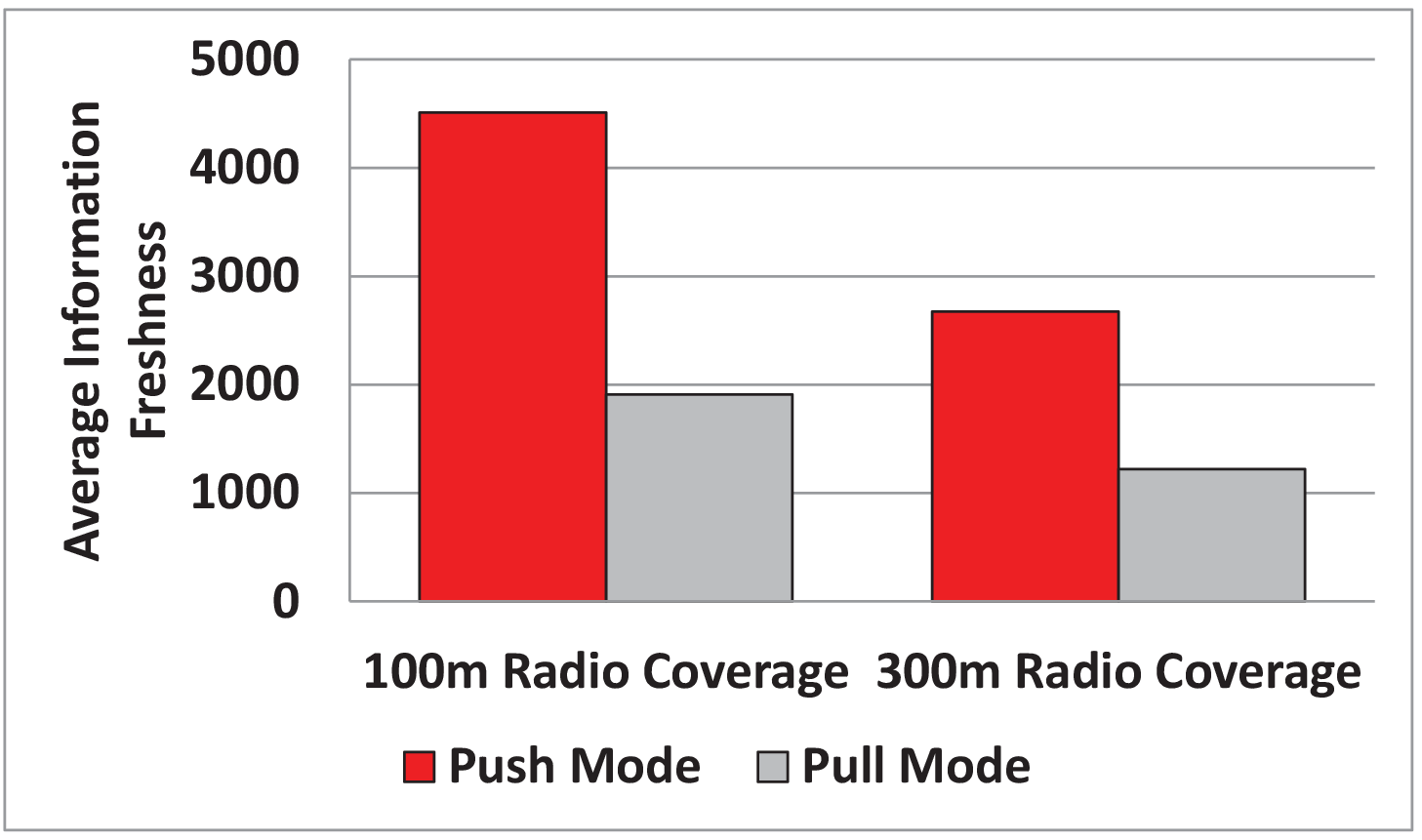}
    }
  \subfigure[Number of Charged EVs]
  {
  \centering
    \label{f44}
    \includegraphics[width=4.1cm,height=2.3cm]{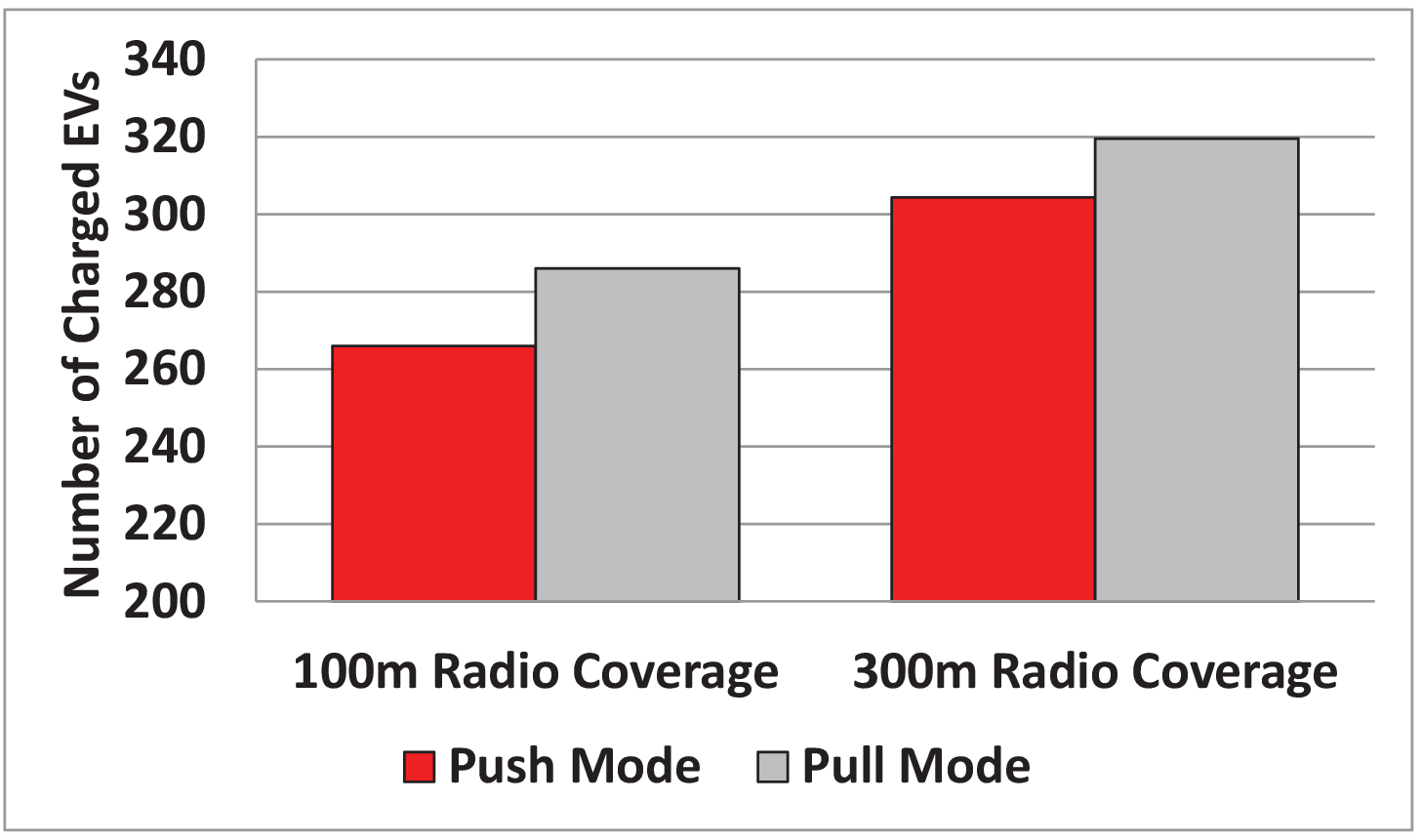}
    }
  \caption{Influence of Radio Coverage}\vspace{-10pt}
\end{figure}

Here, all results are plotted as an average value.
We firstly only vary the radio coverage of all RSUs and maintain that of EVs to evaluate the performance of Push Mode.
Compared to default configuration with 300m radio coverage range, the result in Fig.\ref{f41} shows that the charging system under the Push Mode experiences a longer average waiting time given a short RSU radio coverage.
This is because the chance EVs obtain information from RSUs is significantly reduced in Fig.\ref{f42}, if with a short RSU radio coverage given by 100m.
As such, in Fig.\ref{f43}, the information freshness is also deteriorated in this situation, which leads to a decreased number of charged EVs in Fig.\ref{f44}.
Besides, the influence of only varying EVs' radio coverage is also shown in Fig.\ref{f41}, Fig.\ref{f42}, Fig.\ref{f43} and Fig.\ref{f44} respectively, where the Pull Mode follows the similar trend of Push Mode.

\subsubsection{Influence of Charging Power And EV Speed}
\begin{figure}[htbp]
  \centering
  \subfigure[Average Waiting Time]
  {
  \centering
    \label{f51}
    \includegraphics[width=4.1cm,height=2.3cm]{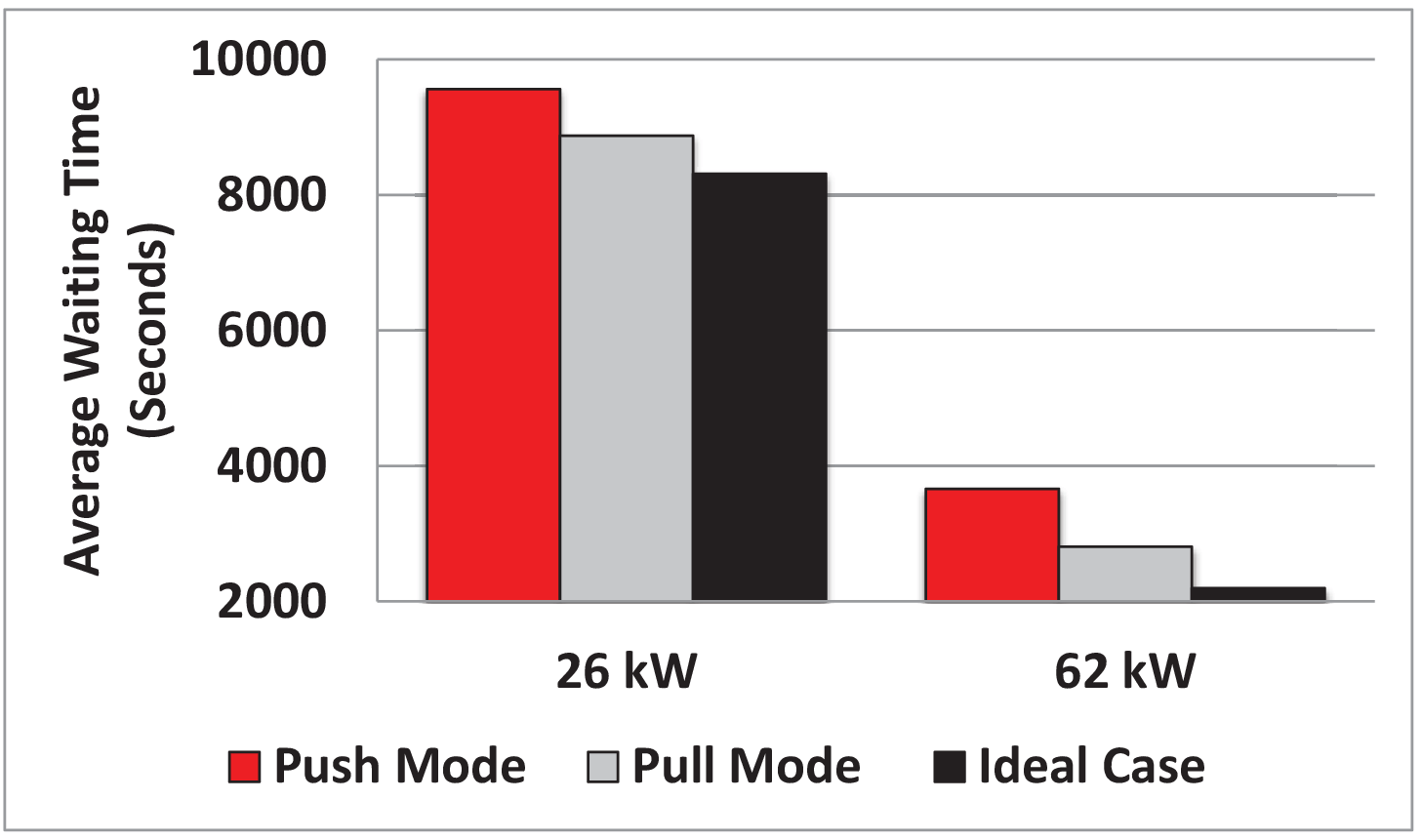}
    }
  \subfigure[Number of Charged EVs]
  {
  \centering
    \label{f52}
    \includegraphics[width=4.1cm,height=2.3cm]{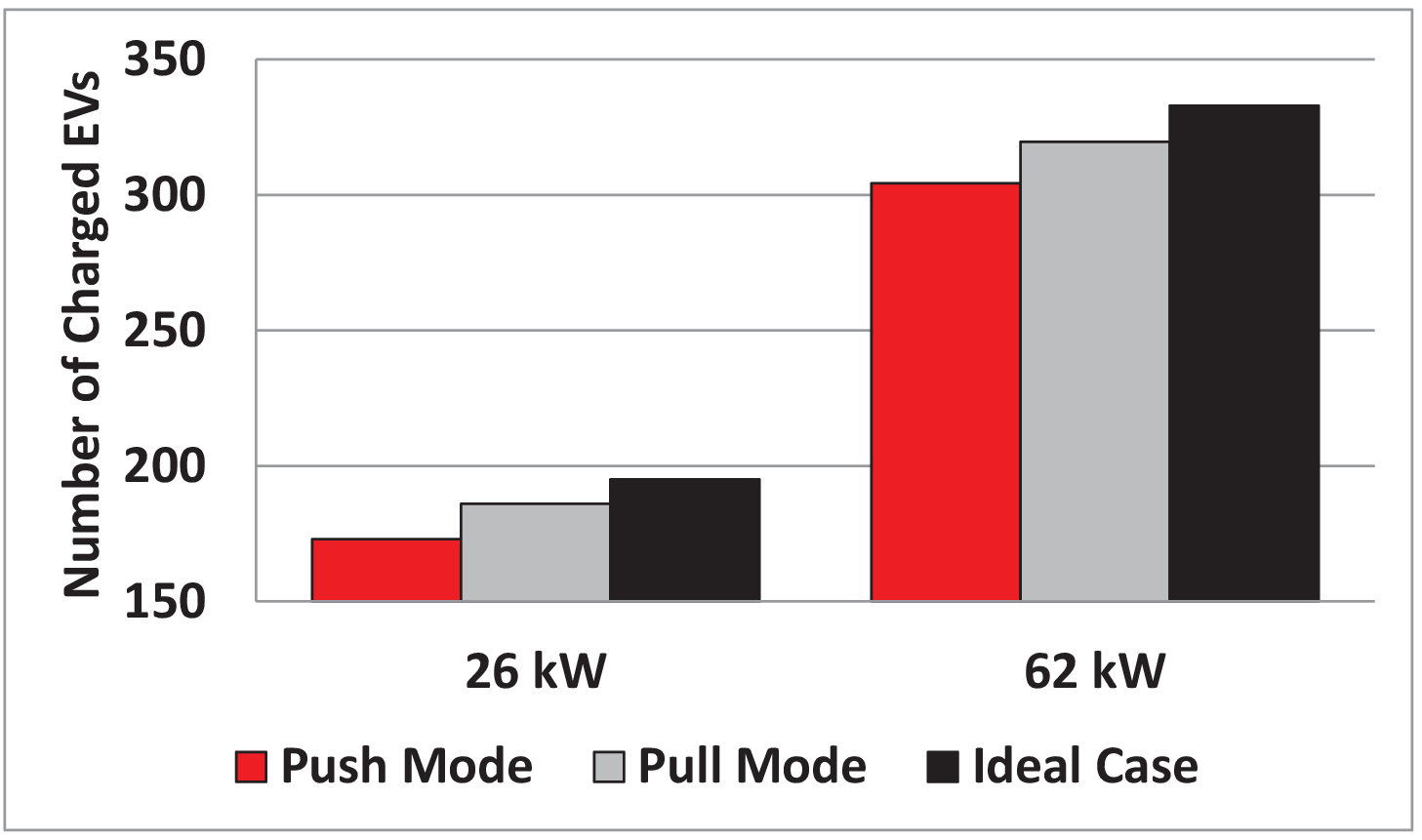}
    }
  \caption{Influence of Charging Power}\vspace{-10pt}
\end{figure}

\begin{figure}[htbp]
  \centering
  \subfigure[Average Waiting Time]
  {
  \centering
    \label{f53}
    \includegraphics[width=4.1cm,height=2.3cm]{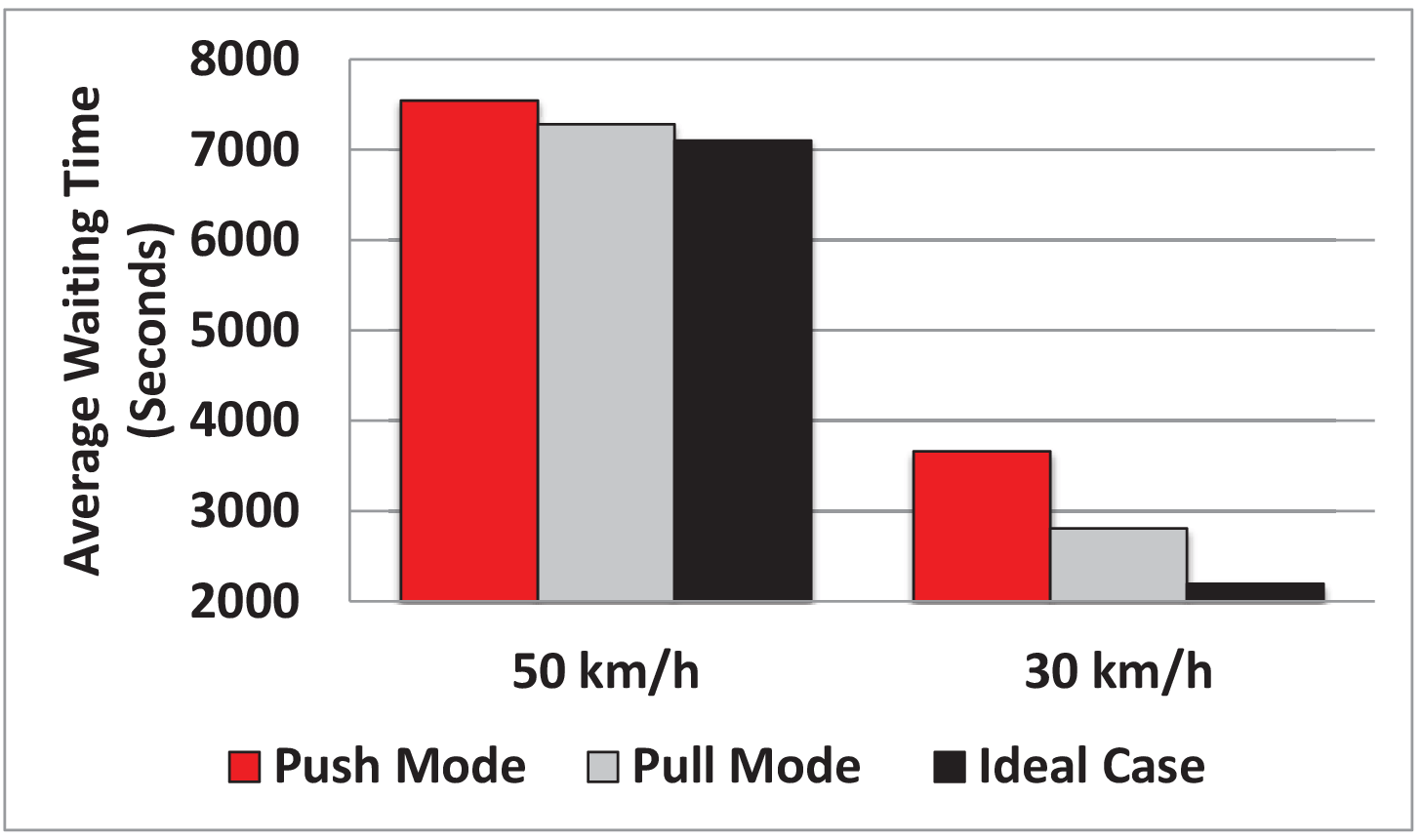}
    }
  \subfigure[Number of Charged EVs]
  {
  \centering
    \label{f54}
    \includegraphics[width=4.1cm,height=2.3cm]{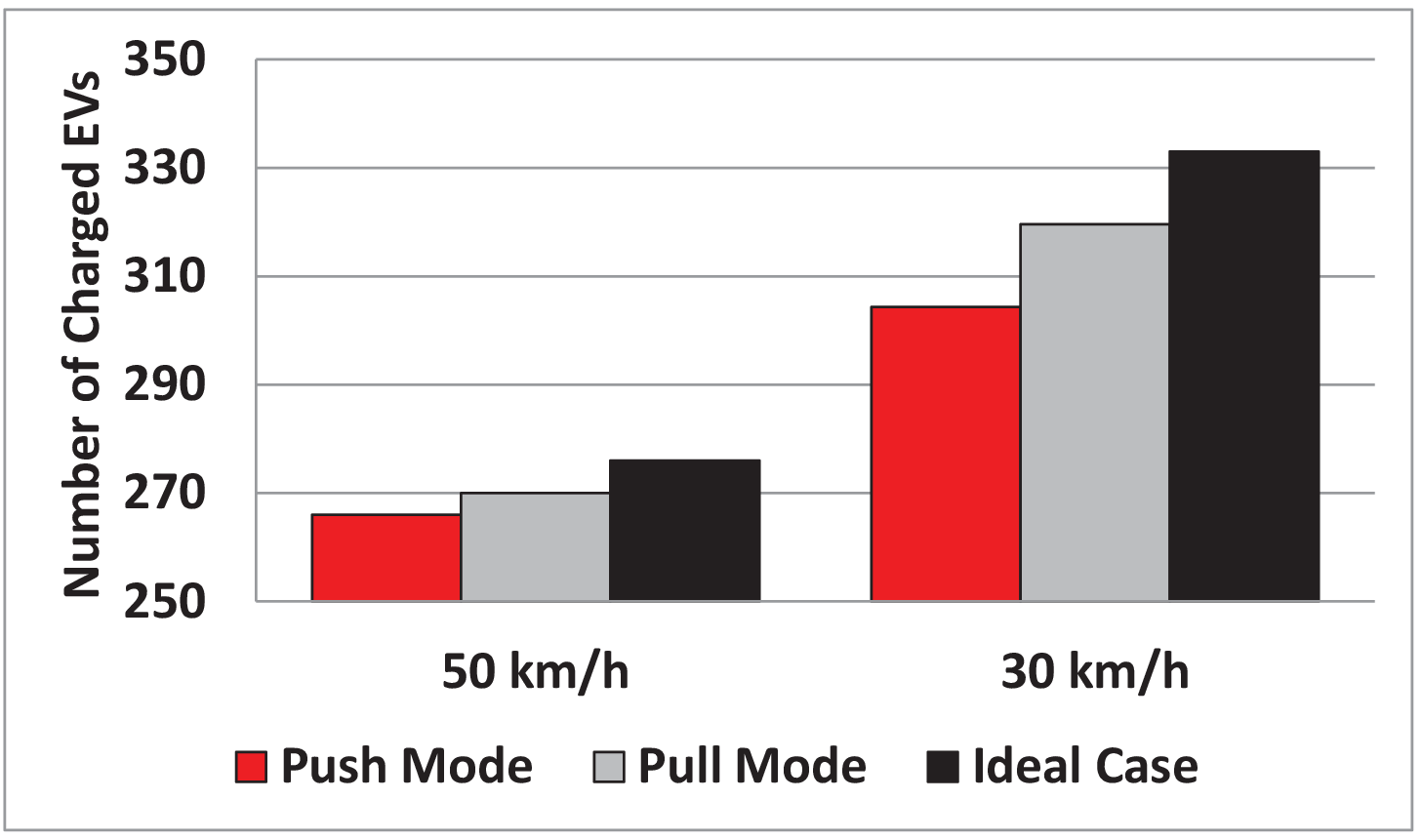}
    }
  \caption{Influence of Minimum EV Speed}\vspace{-10pt}
\end{figure}
Without loss of generality, we provide results in relation to CS charging power (in relation to the service rate) and EV speed (in relation to arrival rate).
Results in Fig.\ref{f51} and Fig.\ref{f52} still follow the performance trend among the Push and Pull Modes as well as Ideal Case, where using faster charging rate improves the average waiting time and number of charged EVs, compared to the case using 26 kW.
In addition, the performance in Fig.\ref{f53} and Fig.\ref{f54} is degraded given the increased minimum EV speed to 50 km/h.
This is because that the charging requests of EVs will become more frequent, as EVs will consume their electricity faster.

\subsubsection{Influence of RSU Number}
\begin{figure}[htbp]
  \centering
  \subfigure[Average Waiting Time]
  {
  \centering
    \label{f61}
    \includegraphics[width=4.1cm,height=2.3cm]{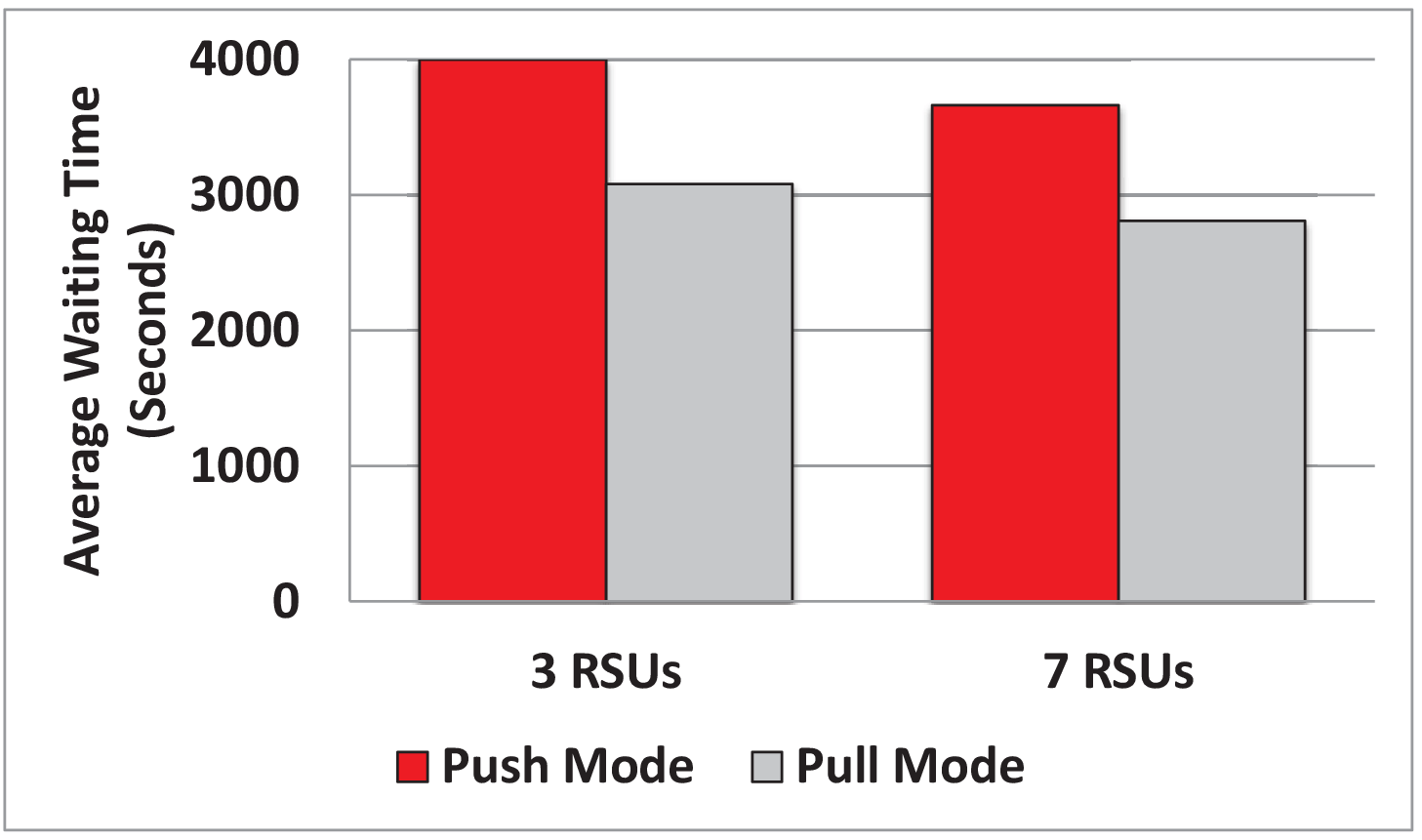}
    }
  \subfigure[Number of Times EVs Obtain Information]
  {
  \centering
    \label{f62}
    \includegraphics[width=4.1cm,height=2.3cm]{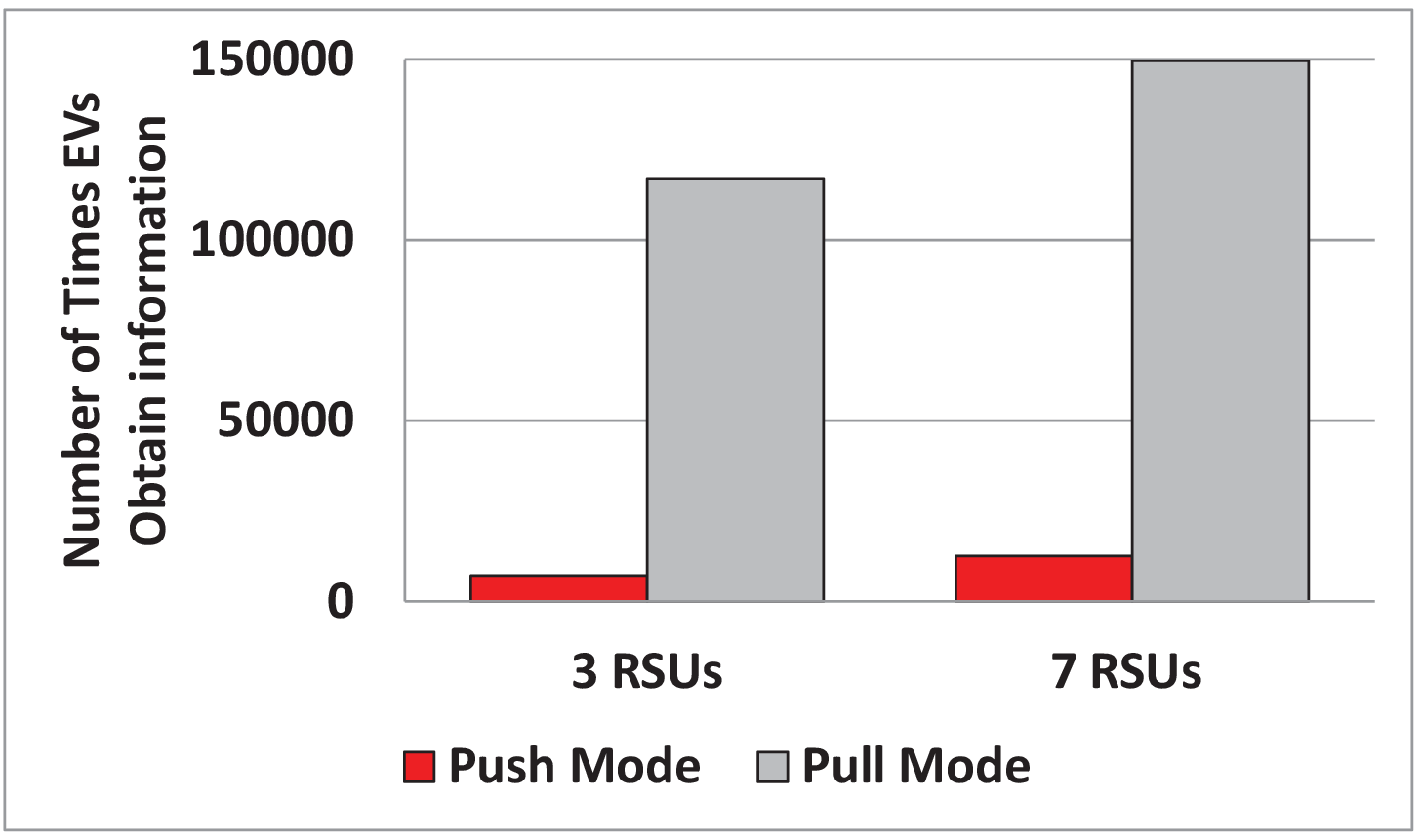}
    }
    \subfigure[Average Information Freshness]
  {
  \centering
    \label{f63}
    \includegraphics[width=4.1cm,height=2.3cm]{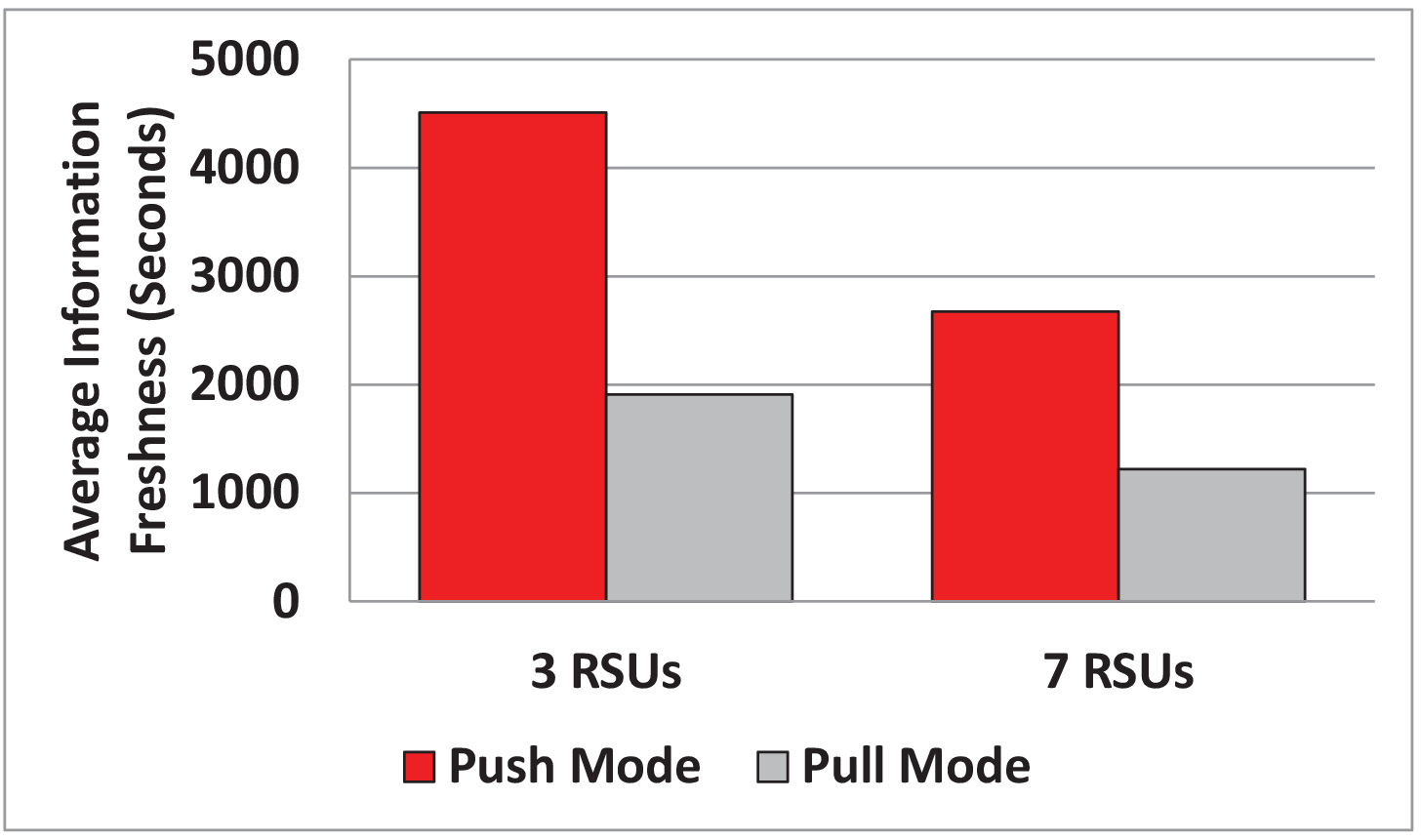}
    }
  \subfigure[Number of Charged EVs]
  {
  \centering
    \label{f64}
    \includegraphics[width=4.1cm,height=2.3cm]{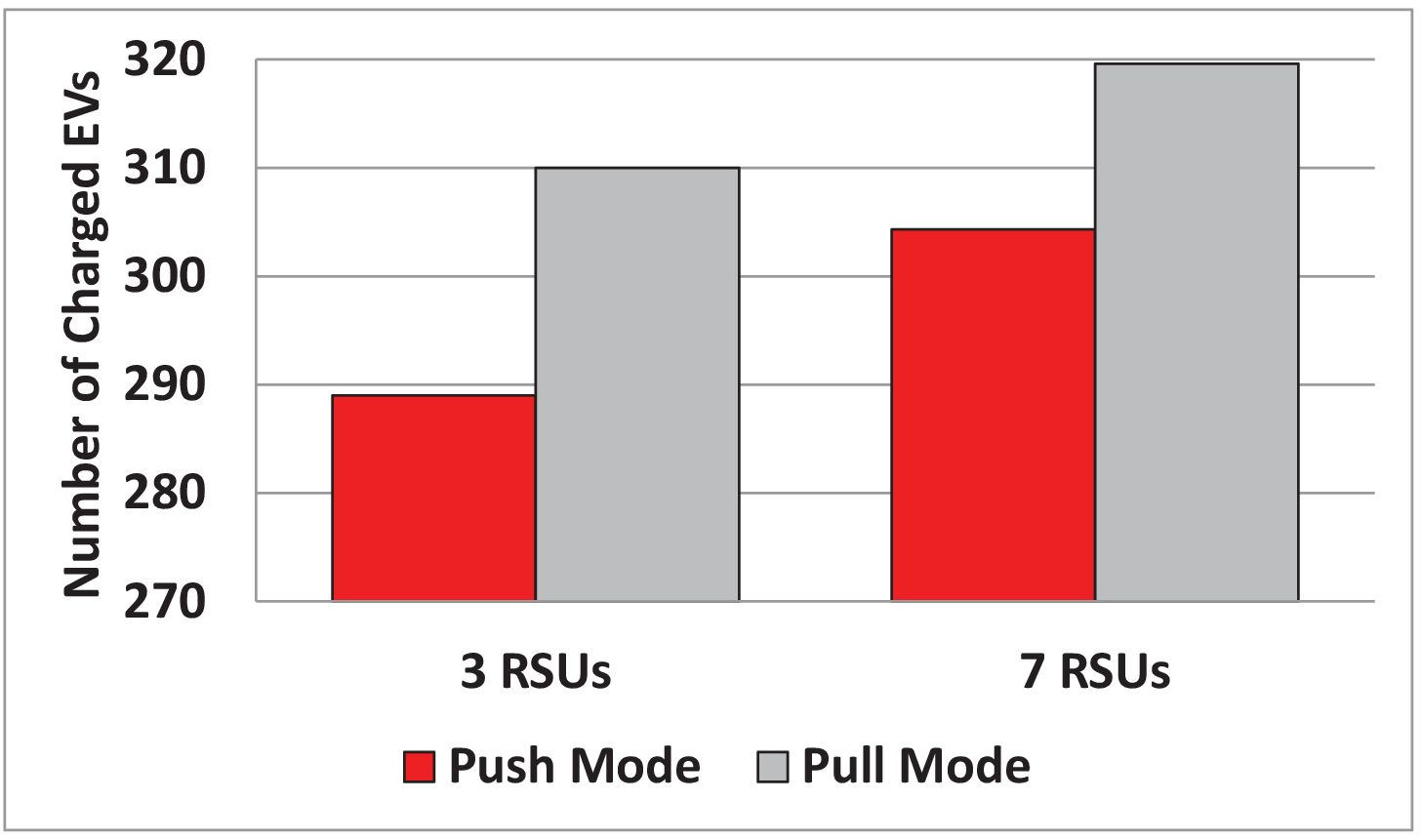}
    }
  \caption{Influence of RSU Density}\vspace{-10pt}
\end{figure}
Here, we remove RSU7, RSU8, RSU10, RSU11 in Fig.\ref{hel} and fix update interval and radio coverage range to be 100s and 300m respectively.
In Fig.\ref{f61}, Fig.\ref{f62}, Fig.\ref{f63} and Fig.\ref{f64}, we observe that reducing the number of RSUs somehow degrades performance, compared to that given 7 RSUs case.

\subsubsection{Discussion on System Scalability}
Key observations from these results could be referred to \cite{iccve}.
Here, the advantage of the proposed system is the scalability that any available RSU in network would continually bridge information disseminated from CS to EV, if some other RSUs fail to work.
This is different from the Ideal Case where the charging management would not be operated if the centralized controller fails.

Although with the advantage in terms of application, a major concern is how frequent the information should be published, as the radio coverage between adjacent RSUs is not ubiquitous.
In the worst case if the publication is extremely infrequent, the EV may not obtain any information from passed RSUs particularly under the Push Mode.
In contrast, a seamless publication is whereas costly as the fresh condition information of CS is only necessary when the EV needs to select a CS.
Here, increasing the update interval to 900s inevitably degrades performance.
In contrast, the performance given 100s update interval (under the Pull Mode) is able to achieve a close performance based on Ideal Case.
This implies that by carefully controlling the CS update interval, an approximately optimal performance could be achieved even without using realtime CS condition information.
In other words, the advantage of such RSU based communication over cellular network communication is concluded as a scalable and cost-efficient system.

Due to decoupling between publishers and subscribers, the end-to-end connections between CSs and EVs are avoided.
Instead, an EV just connects RSU which is close to it.
As a result, we can have scalability (i.e., the number of connections in CS sides does not depend on the number of EVs) and efficiency (i.e., fast connection establishment and reduced bandwidth usage), as the benefits of P/S based communication between CSs and EVs against point-to-point communication.
By modifying the content for information publication, the proposed P/S communication framework can also support the battery replacement scenario \cite{cyber} or the pricing concerning \cite{CitationKey}.
Further to these, since our focus in this article is a proposal of communication framework, deploying RSUs at appropriate places to improve the charging performance is left to our future work.

\section{Proposal of the Advanced Pull Mode With Remote Reservation Service}
In previous section, we have proposed the Push and Pull Modes, using publicly deployed RSUs to bridge the information required for charging.
However, the decision making at EV side only considers the instantaneous queuing time of each CS, without predicting its condition in a near future.
With this in mind, we propose an advanced communication framework based on the Pull Mode, by enabling the EV passing through RSUs to further publish their charging reservation.
Here, the reservation will be bridged by RSUs to the EV's selected CS.
Upon this anticipated information, the decision making based on the minimum expected waiting time at a CS can be estimated considering EVs' future movement. Detailed in subsection \uppercase\expandafter{\romannumeral4.C} and subsection \uppercase\expandafter{\romannumeral4.E}, any EV needs charging service will keep track of the charging time of EVs locally parking at a CS, as well as other EVs with an earlier arrival time heading to this CS. This is different from that in Pull Mode, where EVs only need to know an abstract status about CS.

\subsection{Discussion on Two ETSI Standards for Supporting Our Proposals}
\begin{figure}[htbp]
\begin{center}
\includegraphics[width=8.8cm,height=4cm]{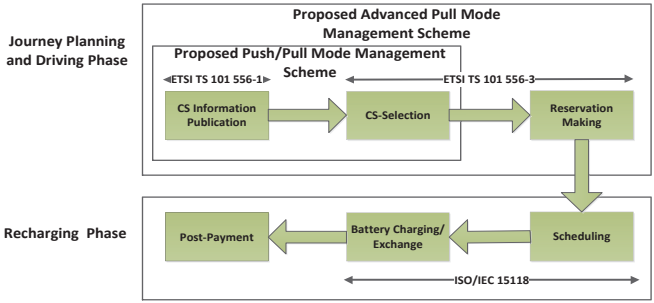}
\caption{The EV Charging Management Cycle}\vspace{-10pt}
\label{cycle}
\end{center}
\end{figure}

In Fig.\ref{cycle}, the CS condition information is published following the format detailed in ``ETSI TS 101 556-1'' standard \cite{etsi1}.
Here, the ``ETSI TS 101 556-1'' standard already includes several ITS entities, e.g., RSU to help to broadcast CSs condition information to EVs, whereas we investigate a topic based P/S mechanism (via either Push/Pull Mode) to efficiently support this communication purpose rather than point-to-point mode.
Upon a CS-selection decision (our technique contributions detailed in Sections \uppercase\expandafter{\romannumeral3} and \uppercase\expandafter{\romannumeral4}) is made at EV side, the EV reservation is published following the format detailed in ``ETSI TS 101 556-3'' standard \cite{etsi3}.
Here, we still rely on RSU to bridge the EV reservation publication to a certain CS.
Note that this CS will further publish its local queuing information together with a number of recorded EVs reservation information following the ``ETSI TS 101 556-1'' standard.

Focusing on ``Journey Planning and Driving Phase'', our proposals on the Push/Pull Mode communication framework to support basic EV charging service as well as the Advanced Pull Mode to support reservation based service are fully compatible with these two standards.
Note that our proposed communication framework can also support a reservation updating operation. Specifically, an EV may publish an update about its reservation, if it can not arrive at the selected CS on time due to traffic congestion, such that its original reservation can be rescheduled. Upon receiving reservation updates due to experienced traffic uncertainties on the EV side, a CS may publish this information periodically to EVs through RSUs.

As discussed in \cite{6687948}, when considering multiple CSs that belong to different grid operators, the market competition between them in attracting EVs for charging can be envisaged. A CS can potentially attract more EVs by publishing with a cheaper electricity price compared to its competitors covering the same region. Of course such price setting will be mainly subject to the energy availability within the grid and the external demand. As a consequence, the price of electricity can be different for CSs belonging to specific operators, even if they are located close to each other. While the difference between the published prices by companies is normally minor, this may still influence decision-making by the EV drivers, in addition to the waiting time factor that is discussed in this paper. On the other hand, the ``ETSI TS 101 556-3" standard further supports pre-payment functions that that can be used by EVs reservation information. In case of pre-payment, the External Identification Means (EIM) is also included. The procedure for obtaining and managing of the External Identification is specific to the local EV charging system. For example, there could be payment cards or virtual tokens sold that include some EIM identifier.

In ``Recharging Phase'', each CS performs scheduling \cite{6919255} for EVs already parking herein based on the FCFS order, or even with smart method by knowing the anticipated EVs arrival information (as included in EVs reservation information). Then, either ordinary battery charging or exchange service will be provided by CS following ISO/IEC 15118 standard \cite{isoiec}, and a payment is posted.

\subsection{Overview of Advanced Pull Mode}
\begin{figure}[htbp]
\begin{center}
\includegraphics[width=9cm,height=5cm]{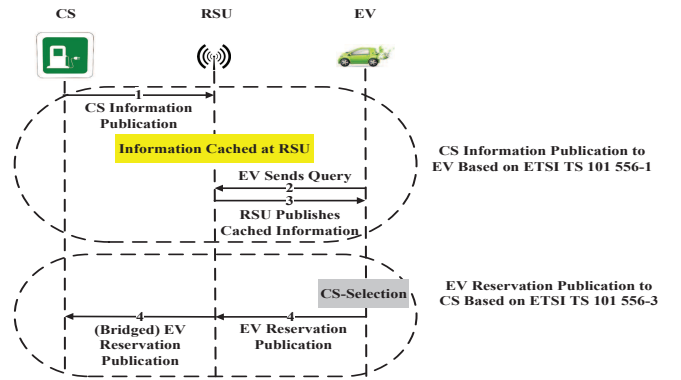}
\caption{Time Sequences for Advanced Pull Mode}\vspace{-10pt}
\label{reservationtimesequence}
\end{center}
\end{figure}

Based on above discussion, the entire timing sequences for on-the-move EV charging under the Advanced Pull Mode (shown in Fig.\ref{reservationtimesequence}) are listed as follows:
\begin{enumerate}
\item An EV accesses CS condition information from RSUs, by referring to the Pull Mode in Section \uppercase\expandafter{\romannumeral3}.
\item Given a low electricity status, the EV selects where to charge using its accessed information.
\item If this EV (which has made decision on where to charge) encounters any RSU on the road, the EV will publish its charging reservation to its selected CS, through the encountered RSU.
\end{enumerate}

\begin{figure}[htbp]
\begin{center}
\includegraphics[width=8.3cm,height=4.8cm]{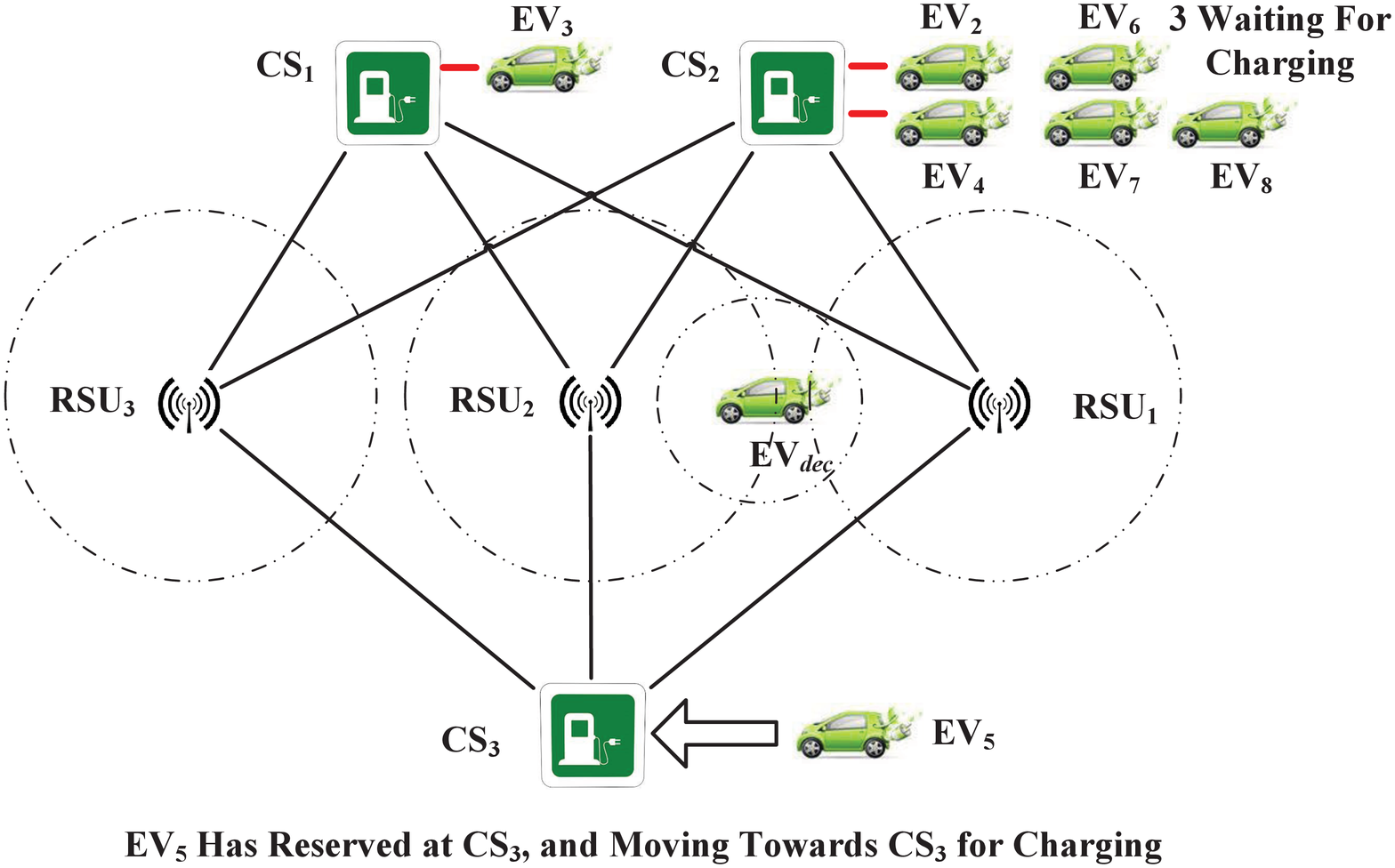}
\caption{An Overview of Remote Reservation Service}
\label{reservationsystem}
\end{center}
\end{figure}

As an example shown in Fig.\ref{reservationsystem}, the EV needs to select a CS for charging, namely EV$_{dec}$, recently passed through and learnt from RSU$_1$ that there were 5 EVs parking at CS$_2$ and 1 EV charging at CS$_1$.
Meanwhile, it also learnt from RSU$_1$ that EV$_5$ had reserved CS$_3$ for charging and would take 20 minutes to reach CS$_3$.
Assuming EV$_{dec}$ needs 30 minutes to reach CS$_3$, it realizes that a potential waiting time at CS$_3$.
As such, the decision to select a less loaded CS is made based on the historical information in relation to the waiting time at CSs and other EVs' reservation.
Later on, EV$_{dec}$ will publish its CS-selection decision through RSU$_2$ passed by, including when it will arrive at its selected CS as well as the charging time it will require upon that arrival.

To the best of our knowledge, only previous work \cite{Qin:2011:CSM:2030698.2030706} as reviewed in Section \uppercase\expandafter{\romannumeral2} considers EVs' future movement to select CS for charging.
In detail, our proposal has the following substantial differences compared to that work:

1) This previous work assumes the spatially distributed CSs are deployed at highway, such that EVs will pass through CSs during their journeys.
In contrast, our scenario is more flexible that the mobility of EVs are without any restriction, and can travel towards any CS geographically deployed at certain place for charging.
Therefore, that previous work is not applicable in our scenario.

2) Another limitation of this previous work is that the number of charging slots is not considered for calculating the expected waiting time at a CS.
Here, the charging procedure is parallel among multiple charging slots, different from the case where the time to await charging is linearly reduced if using single charging slot.

3) Since the decision making in our communication framework is at the EV side, our proposal processes each  EVs' reservation information concerning privacy issue where the CS-selection decision uses the anonymous EV information. This is different from that previous work in which the decision is made at CS side using EVs' reservation information, particularly the EV ID is not hidden through communication.

\subsection{Estimating CS's Available Charging Time}
In order to determine how long an arrival EV will wait for charging, it is essential to estimate when a charging slot is available for charging, considering a number of EVs have already parked at a CS.
If using only one charging slot, the available time for charging is linearly reduced with a constant rate, depending on the CS charging power.
This is because an arrival EV has to wait until all EVs in front of the queue have finished their charging.
However, the situation is not applicable if using multiple charging slots, because the charging for a number of EVs is in parallel.

Based on notations in TABLE \ref{configure1}, here we consider two types of queues respectively.
Those EVs which are under charging are characterized in the queue of $N_C$, while those still waiting for charging are characterized in the queue of $N_W$.
\begin{itemize}
\item Presented between lines 2 and 5 in Algorithm \ref{print}, if none of EVs is under charging, the current time in network, as denoted by $T_{cur}$, is estimated as the available time for charging per charging slot.
    In special case that if charging slots are not fully occupied, the outputs include the charging finish time of each EV currently under charging (as calculated by aggregating its charging time and $T_{cur}$), and $T_{cur}$ for each free charging slot.
\item Alternatively, those EVs in the queue of $N_W$ will be sorted based on the FCFS order, following the previously stated scheduling method.
    As presented between lines 13 and 14, they will be inserted into the queue of $N_C$, once a charging slot is free depending on the minimum charging time of the EV occupied this slot.
    This charging time is calculated based on Algorithm \ref{alg1}, as the minimum charging time of those EVs under charging.
    Meanwhile, presented between lines 10 and 11, the charging time of other EVs (not with the minimum charging time) still under charging are reduced by this value until the charging is finished, considering the parallel charging procedure among multiple charging slots.
\item The above loop operation ends when all EVs in the queue of $N_W$ have been inserted into the queue of $N_C$.
Starting from line 18, then the available charging time for each slot are estimated, by aggregating the charging time of those EVs which occupy the charging slots and the current time in network $T_{cur}$, presented at line 19.
\end{itemize}
This output together with the instantaneous queuing time that is calculated via Algorithm \ref{alg2}, are published as the local condition information of a CS within each publication interval. Note that any information related to CS is public, thus there is no privacy concern.
\begin{algorithm}[htbp]\footnotesize
\caption{Estimate CS's Available Charging Time}
\label{print}
\begin{algorithmic}[1]
\STATE define LIST $/\ast$The information publication containing available time for charging per charging slot$\ast/$
\IF {$(N_C<\vartheta)$}
\STATE adopt $T_{cur}$ for each free charging slot, added in LIST
\STATE aggregate the charging time of EV under charging and $T_{cur}$ for each occupied charging slot, added in LIST
\ENDIF
\STATE sort the queue of $N_W$ according to FCFS
\FOR{$(i=1$; $i\leq N_W$; $i++)$}
\STATE define $\text{VALUE}=\text{Output From Algorithm \ref{alg1}}$
\FOR{$(j=1$; $j\leq N_C$; $j++)$}
\IF {$\left(\frac{E^{max}_{ev_{(j)}}-E^{cur}_{ev_{(j)}}}{\beta}\neq \text{VALUE}\right)$}
\STATE update the charging time of EV$_j$ to be $\left(\frac{E^{max}_{ev_{(j)}}-E^{cur}_{ev_{(j)}}}{\beta}-\text{VALUE}\right)$
\ELSE
\STATE delete EV$_j$ from the queue of $N_C$
\STATE insert EV$_i$ into the queue of $N_C$
\ENDIF
\ENDFOR
\ENDFOR
\FOR{$(j=1$; $j\leq N_C$; $j++)$}
\STATE add $\left(\frac{E^{max}_{ev_{(j)}}-E^{cur}_{ev_{(j)}}}{\beta}+T_{cur}\right)$ into LIST
\ENDFOR
\RETURN LIST
\end{algorithmic}
\end{algorithm}

\subsection{Publishing EV's Reservation Information}
As introduced previously, the core of remote reservation service is to predict the expected waiting time at a CS using EVs' reservation information.
When EV$_{dec}$ requires charging, it will utilize the received other EVs' reservation information to estimate the expected waiting time at a CS in the near future.
For this purpose, any EV which has made CS-selection decision and is already travelling towards its selected CS, will further publish the following reservation information in relation to that CS through the communication between a RSU.
\begin{itemize}
\item Based on the travelling time $T^{tra}_{ev}$ calculated from the current location of EV to that CS via the shortest road path, the expected arrival time $T^{arr}_{ev}$ is given by:
\begin{equation}\footnotesize
T^{arr}_{ev}=T_{cur}+T^{tra}_{ev}
\end{equation}

\item Besides, we denote $T^{cha}_{ev}$ as the expected charging time upon that arrival, where:
\begin{equation}\footnotesize
T^{cha}_{ev}=\frac{E^{max}_{ev}-E^{cur}_{ev}+S_{ev}\times T^{tra}_{ev}\times\alpha}{\beta}
\end{equation}
Here, $(S_{ev}\times T^{tra}_{ev}\times\alpha)$ is the energy consumed for movement travelling to the selected CS, based on a constant $\alpha$ measuring the energy consumption per meter.
\end{itemize}

Since the decision is made at EV$_{dec}$ side, there is a privacy concern if releasing the IDs of other EVs to EV$_{dec}$.
Motivated by this concern, each CS will integrate the information in relation to a number of EVs which reserve at here for charging, with the CS local information for publication.
As observed from the format in TABLE \ref{format}, the IDs of those EVs reserving for charging at CS are hidden.
As such, EV$_{dec}$ will not obtain any knowledge about who else has reserved for charging, since only a list of entries containing the arrival time and reserved charging time upon that arrival are received.

In addition, the published arrival time does not disclose the EV's location.
This is because this arrival time is estimated depending on the location of EV and its corresponding speed at that time, whereas these two information will not be released through any communication.
In this context, EV$_{dec}$ will also not obtain any knowledge about the locations of other EVs.

One concern of such reservation reporting is that it is inherently entitled with a possibility to introduce system instability through service attacks against CSs, since EVs send reservation information to CSs for future charging.
The assumption that reservation information is trustworthy is vulnerable without ensuring the integrity of messages from EVs to CSs on end-to-end aspects.
E.g., forged or wrong reservation information are continuously delivered to CSs through RSUs,  CSs will compute quite imprecise estimation for charging available time and advertise imprecise information to EVs through RSUs.
The general secured vehicular communication framework in \cite{4689253} can be applied to enable secured delivery of EV reservation requests towards CSs. Due to the space limit we will not provide detailed information on this enabling technique in this article.

\begin{table}[htbp]\scriptsize
\renewcommand{\arraystretch}{1.3}
\caption{Format of Published Information From CS Side}
\label{format}
\centering
\begin{tabular}{|c|c|c|}
\hline
\multicolumn{3}{|c|}{CS ID}\\\hline
\multicolumn{3}{|c|}{CS$_{3}$}\\\hline\hline
\multicolumn{3}{|c|}{Instantaneous Queuing Time}\\\hline
\multicolumn{3}{|c|}{3060s}\\\hline\hline
\multicolumn{3}{|c|}{Available Charging Time Per Charging Slot}\\\hline
\multicolumn{3}{|c|}{[3300s, 3950s, 4210s]}\\\hline\hline
\multicolumn{3}{|c|}{Reservation Information}\\\hline
Reservation Entry&Arrival Time &Reserved Charging Time \\\hline
1&3500s &730s\\\hline
2&4700s&700s\\\hline
\end{tabular}
\end{table}

\subsection{Estimating EV's Expected Waiting Time}
\begin{table}[htbp]\scriptsize
\caption{Structure For Maintaining CS Information, at RSU and EV Side}\vspace{-10pt}
\label{ev}
\centering
\begin{tabular}{|p{0.5cm}|p{2.2cm}|p{2.2cm}|p{2.2cm}|}
\hline
{Key} &\multicolumn{3}{|c|}{Value}\\
\cline{1-4}
ID & Instantaneous Queuing Time  at CS & Available Time For Charging Per Charging Slot&Reservation Information of CS \\\hline
CS$_{3}$ & 3060s & [3300s, 3950s, 4210s] &[3500s, 730s], [4700s, 700s]\\\hline
\end{tabular}
\end{table}
Both RSU and EV will build a ``Map $<$Key, Value$>$'' structure following TABLE \ref{ev}.
The ``Key'' is the entry for each CS, while the ``Value'' is a tuple consisting of the local information about this CS and those EVs reserving for charging at here.
Whenever a new information is received from RSU, the old information in ``Value'' will be replaced.
The CS-selection decision is then made at EV side using these recorded information.

Upon the received information, the expected waiting time at a given CS can be estimated by EV$_{dec}$.
In special case that if none of the reservation is available, only the instantaneous queuing time at CS is adopted instead.
The detail is presented in Algorithm \ref{reservation}, where $N_R$ stands for the number of entries for the ``Arrival Time, Reserved Charging Time'' pair.
Here, $T^{arr}_i$ and $T^{cha}_i$ are denoted as the arrival time and corresponding expected charging time at $i^{th}$ entry.

\begin{algorithm}[htbp]\footnotesize
\caption{Estimate EV's Expected Waiting Time}
\label{reservation}
\begin{algorithmic}[1]
\STATE define $\text{CFT}=0$ $/\ast$Charging Finish Time$\ast/$
\STATE sort the queue of $N_R$ according to FCFS
\STATE define TEMLIST = Output From Algorithm \ref{print}
\STATE sort TEMLIST with ascending order $/\ast$The Fast Available Time for Charging is at the Head$\ast/$
\FOR{$(i=1$; $i\leq N_R$; $i++)$}
\IF {($T^{arr}_i<T^{arr}_{ev_{(dec)}})$}

    \STATE define $\text{FATC}=\text{TEMLIST.GET(0)}$ $/\ast$Fast Available Time for Charging$\ast/$

    \IF {$(\text{FATC}>T^{arr}_i)$}
        \STATE $\text{CFT}=\text{FATC}+T^{cha}_i$
    \ELSE
        \STATE $\text{CFT}=T^{arr}_i+T^{cha}_i$
    \ENDIF
    \STATE replace TEMLIST.GET(0) with CFT, in TEMLIST
    \STATE sort TEMLIST with ascending order
\ENDIF
\ENDFOR

\IF {$(\text{TEMLIST.GET(0)}>T^{arr}_{ev_{(dec)}})$}
\RETURN $\text{TEMLIST.GET(0)}-T^{arr}_{ev_{(dec)}}$
\ELSE
\RETURN 0
\ENDIF
\end{algorithmic}
\end{algorithm}

The Algorithm \ref{reservation} initially sorts the queue of $N_R$ following FCFS policy, as their charging will be scheduled via this order.
For each $T^{arr}_i$ earlier than $T^{arr}_{ev_{(dec)}}$, as the arrival time of EV$_{dec}$, the former will involve the dynamic update of TEMLIST as returned by Algorithm \ref{print}. The purpose is to estimate when a charging slot will be available for charging upon the arrival of EV$_{dec}$. The TEMLIST is also sorted according to the ascending order such that the fast available time for charging is at the head. In this article, we denote TEMLIST.GET(0) as the first value in TEMLIST.
\begin{itemize}
\item If $T^{arr}_i$ is earlier than the fast available time for charging as released from a charging slot, the time slot about when the charging upon this arrival will be finished, is calculated by aggregating this available time for charging and the corresponding expected charging time $T^{cha}_i$, following line 9.

\item In contrast, the charging finish time is calculated by aggregating $T^{arr}_i$ and $T^{cha}_i$ following line 11.
    This is because a charging slot has already been free given $T^{arr}_i$.
\end{itemize}
By replacing the fast available time for charging with this charging finish time, the available time for charging per charging slot is dynamically updated, until all arrival times in the queue of $N_R$ have been checked for this loop operation.
Note that the TEMLIST will be sorted with ascending order after the process of each arrival time in the queue of $N_R$, such that the fast available time for charging is always at the head of this list for further calculation.

Then the arrival time of EV$_{dec}$ will be compared with the fast available time for charging, as the head value in TEMLIST.
Their differential is estimated as the expected waiting time if EV$_{dec}$ travels towards this CS for charging, as presented between lines 17 and 21.
By recursing Algorithm \ref{reservation} for each CS, the final CS-selection decision is to find the CS with the minimum value of such expected waiting time.

\subsection{Performance Evaluation}

\begin{figure*}[htbp]
\centering
\subfigure[Average Waiting Time]
{
\centering
\label{f71}
\includegraphics[width=5.5cm,height=2.8cm]{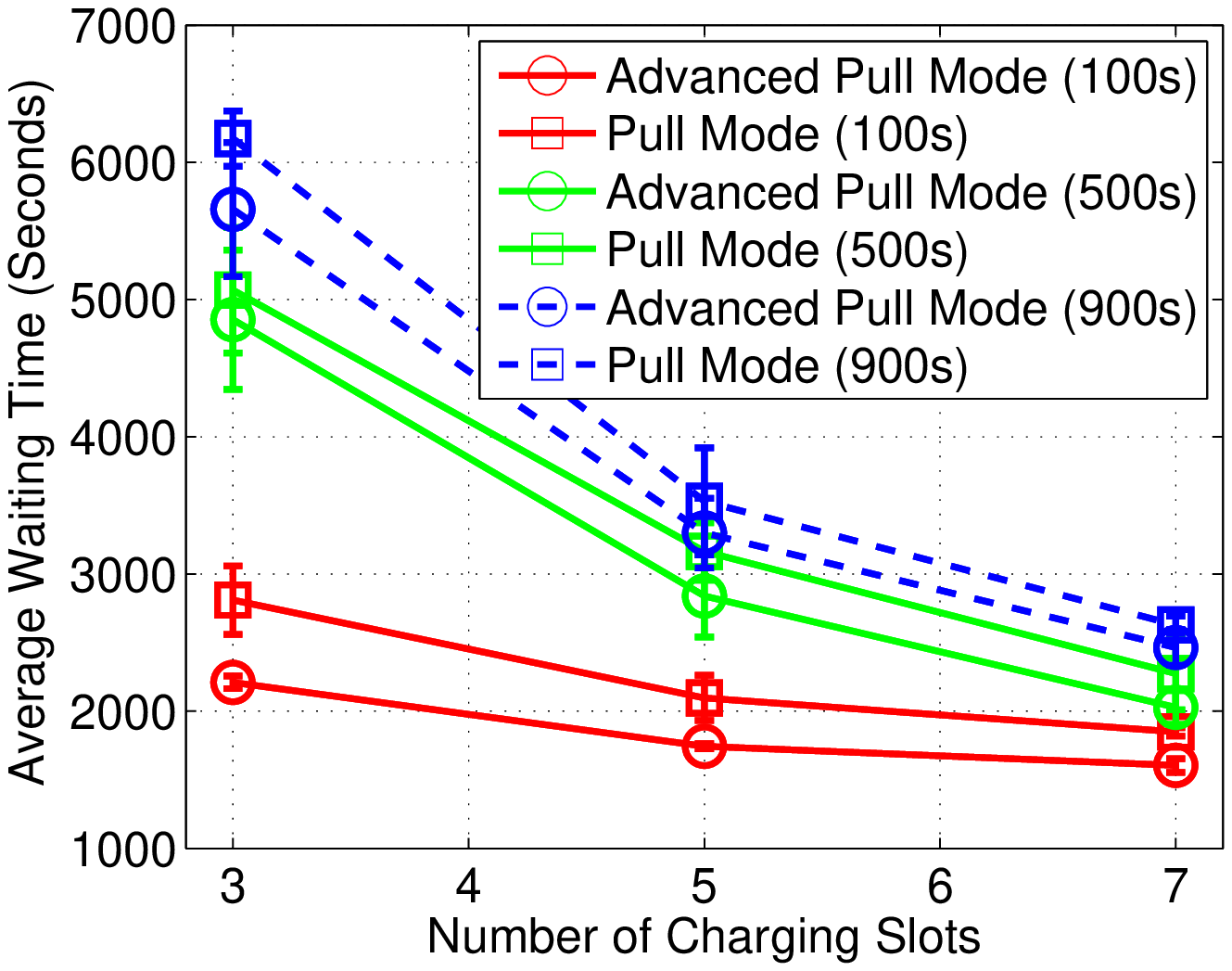}
}
\subfigure[Number of Charged EVs]
{
\centering
\label{f72}
\includegraphics[width=5.5cm,height=2.8cm]{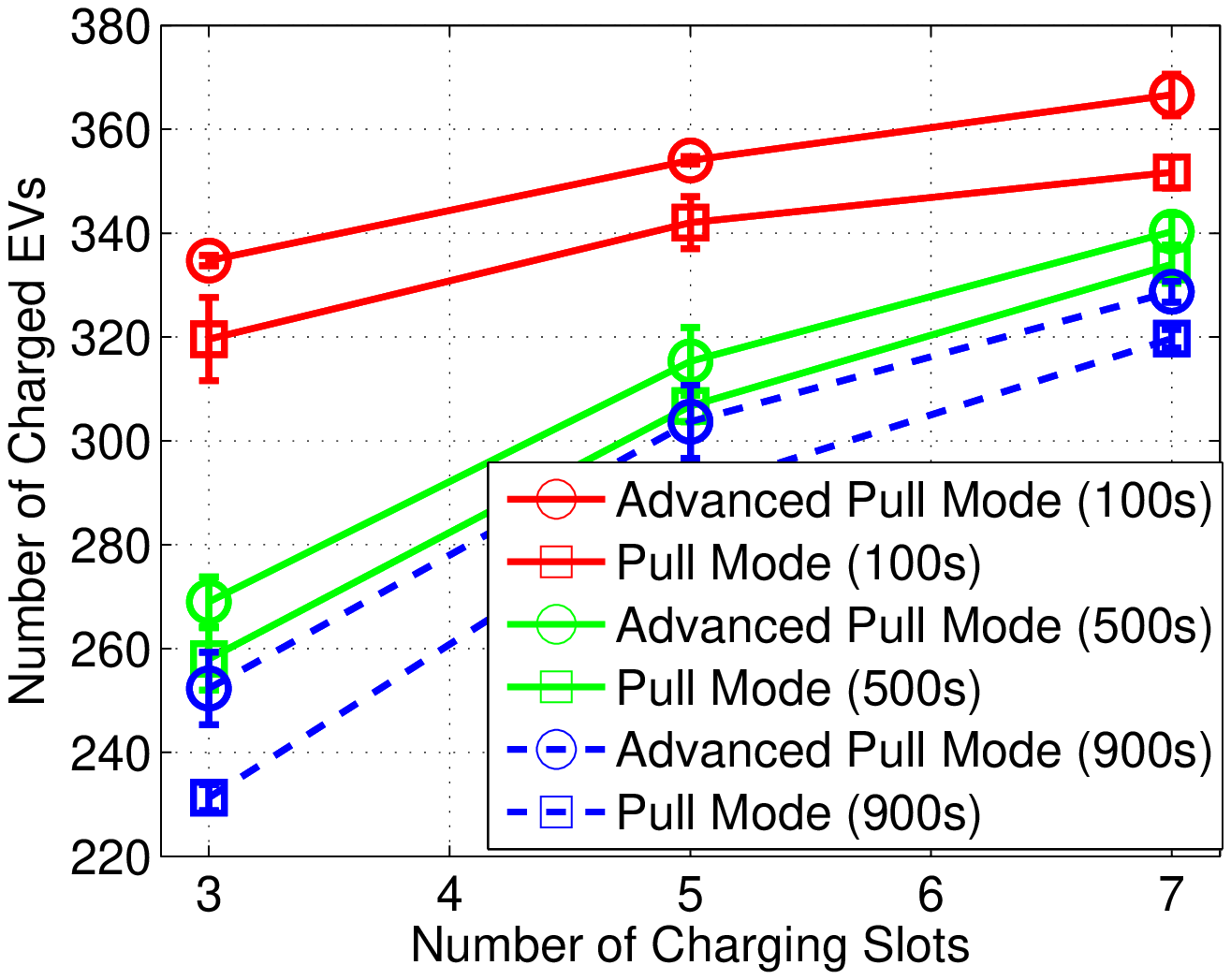}
}
\subfigure[Utilization of CSs (3 Charging Slots)]
{
\centering
\label{f73}
\includegraphics[width=5.5cm,height=2.8cm]{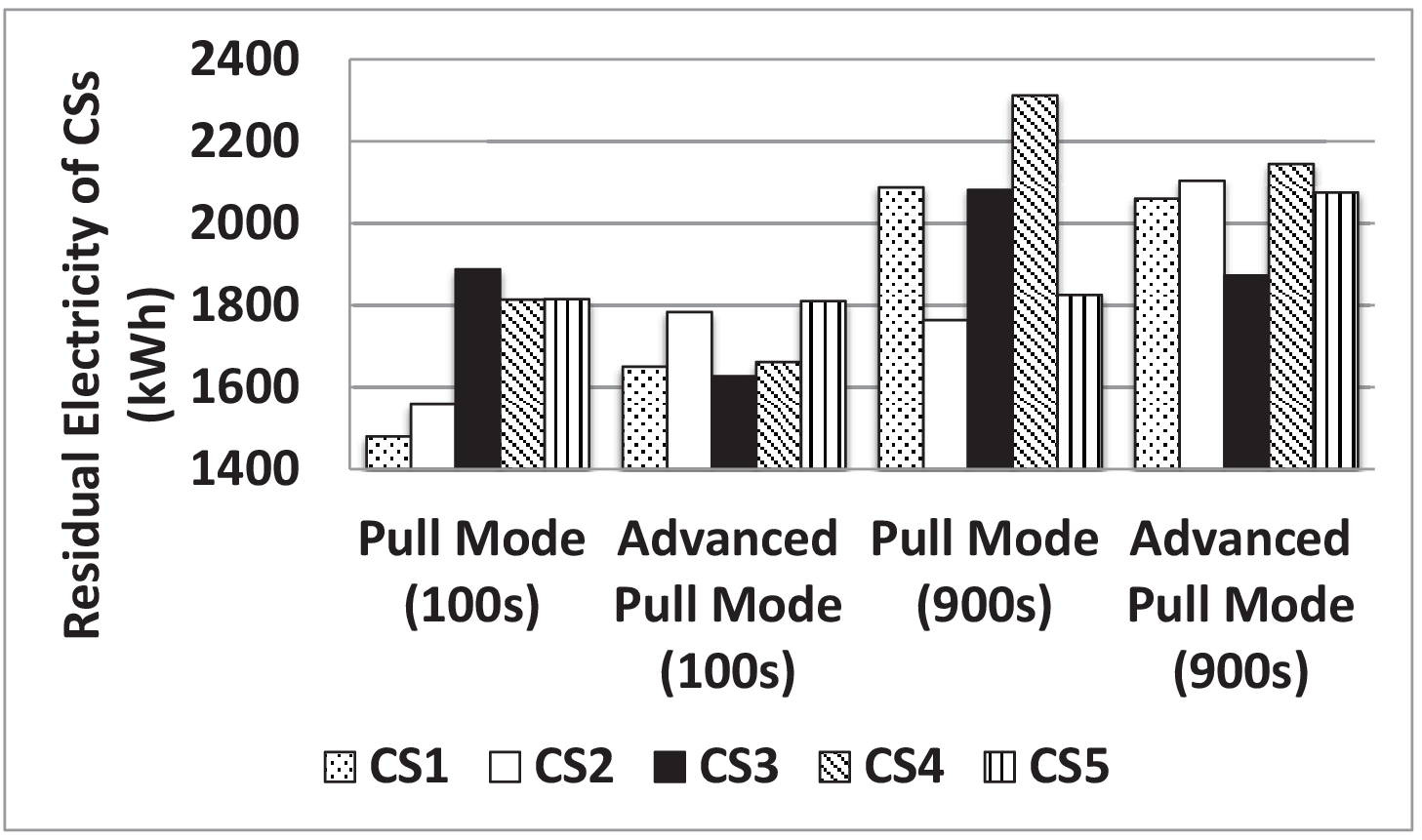}
}
\caption{Comparison Between Pull And Advanced Pull Modes}\vspace{-10pt}
\end{figure*}

The performance evaluation is based on the same scenario as described in Section \uppercase\expandafter{\romannumeral3}.
In Fig.\ref{f71}, we observe increasing the number of charging slots reduces the average waiting time, since the parallel charging process enables more EVs can be charged simultaneously.
Here, both the Pull Mode and Advanced Pull Mode achieve the best performance given 100s update interval, compared to that given 900s update interval.
This is because that a more frequent information publication improves the information freshness at EV side to make accurate CS-selection decision, in particular the reservation in relation to a CS as well as its local information are received with a more recent value.
Compared to the original Pull Mode by only using CS instantaneous queuing information, using EVs' reservation information improves the average waiting time, by considering EVs' future movement to select the CS with the minimum expected waiting time.
Of course, applying more charging slots improves performance for both of them.
The observation in Fig.\ref{f72} shows that the Advanced Pull Mode charges more number of EVs than that under original Pull Mode.
Note that a more fresh information is beneficial to make accurate decision, as such the performance given 100s update interval achieves the highest value.
Concerning energy aspect, the utilization of CSs under these two modes does not differ too much in Fig.\ref{f73}, in case of 100s update interval.
This is mainly due to making CS-selection decision based on the frequently published information.
However, the situation is changed in case of 900s update interval, as the CS-selection decision made via outdated information results in unbalanced number of EVs distributed among CSs.

\section{Conclusion}
In this article, we proposed an efficient communication framework for EV application, based on the P/S mechanism
and deployed RSUs to disseminate the status information of CSs.
Here, two communication modes are specified.
Based on the analysis regarding these two modes, we further developed the entire charging decision making system via ONE simulator for a typical EV scenario.
Results showed that the Pull Mode achieves a better performance regarding shorter waiting time as well as energy balance.
We further propose the Advanced Pull Mode which enables EVs to publish their reservation information in relation to selected CSs.
In this context, EVs benefit from this anticipated information to make smart CS-select decision, which further reduces the EVs' waiting time for charging as well as improving the utilization among CSs.
\bibliographystyle{IEEEtran}
\bibliography{refer}

\begin{IEEEbiography}[{\includegraphics[width=1in,height=1.25in,clip,keepaspectratio]{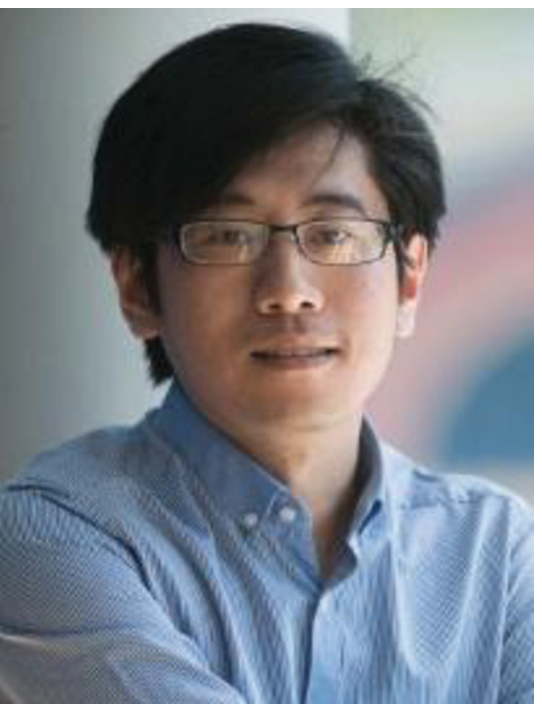}}]{Yue Cao} joined the Institute for Communication Systems (ICS) formerly known as Centre for Communication Systems Research (CCSR), at University of Surrey, Guildford, UK in 2009 and obtained his PhD degree in 2013. He is currently a Research Fellow at the ICS. His research interests focus on Delay/Disruption Tolerant Networks, Electric Vehicle (EV) communication, Information Centric Networking (ICN) and traffic offloading for cellular systems.
\end{IEEEbiography}

\begin{IEEEbiography}[{\includegraphics[width=1in,height=1.25in,clip,keepaspectratio]{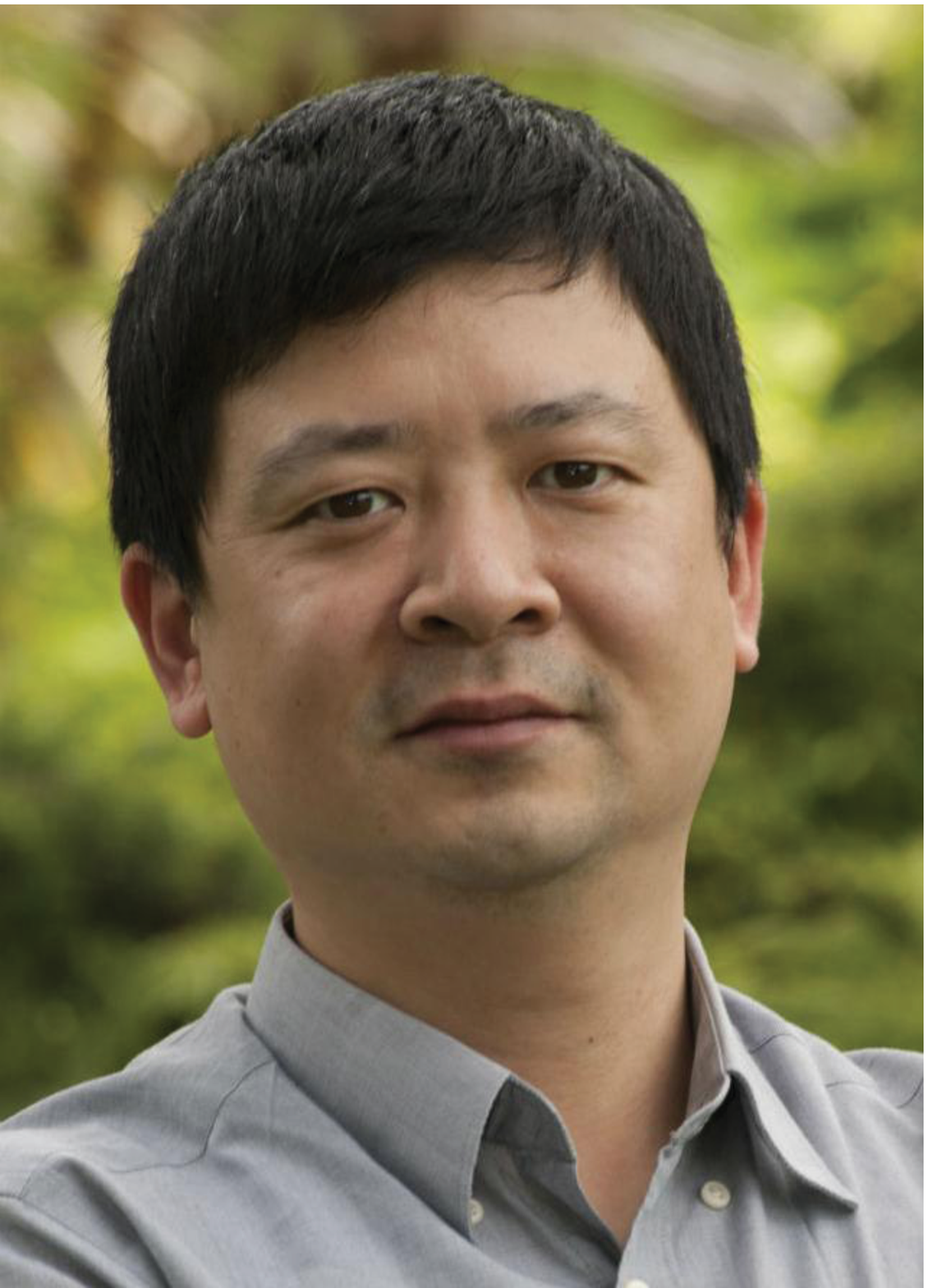}}]{Ning Wang} received his PhD degree from the Institute for Communication Systems (ICS) formerly known as Centre for Communication Systems Research (CCSR), at University of Surrey, Guildford, UK in 2004.
He is currently a Reader at the ICS and his research interests mainly include energy-efficient networks, network resource management, Information Centric Networking (ICN) and QoS mechanisms.
\end{IEEEbiography}

\begin{IEEEbiography}[{\includegraphics[width=1in,height=1.25in,clip,keepaspectratio]{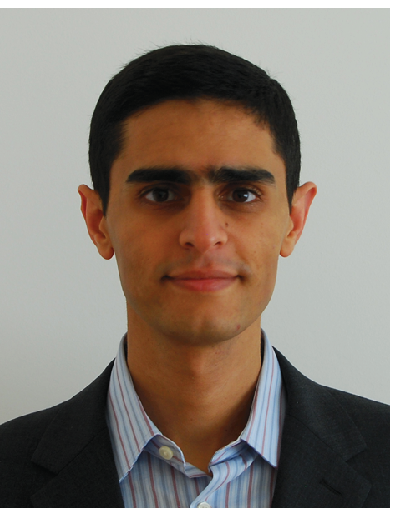}}]{George Kamel} is a Research Fellow in the Institute for Communication Systems (ICS) formerly known as Centre for Communication Systems Research (CCSR), at the University of Surrey, Guildford, UK. He obtained his PhD degree in Telecommunications Engineering from King¡¯s College London, UK in 2010. His current research interests include Information Centric Networking (ICN) and traffic offloading for cellular systems.
\end{IEEEbiography}

\begin{IEEEbiography}[{\includegraphics[width=1in,height=1.25in,clip,keepaspectratio]{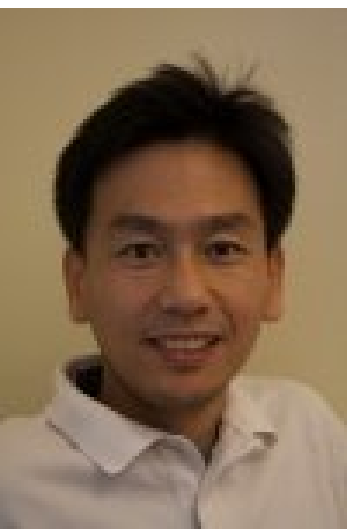}}]{Young-Jin Kim} has been a Member of Technical Staff of network algorithm, routing, and security department at Bell-Labs, Alcatel-Lucent, Murray Hill, NJ since 2010. Dr Kim received his PhD degree in Computer Science from University of Southern California in 2008. His research interests include software platforms, algorithms, and protocols in Software-Defined Networking (SDN), Information Centric Networking (ICN), and cloud network infrastructures.
\end{IEEEbiography}

\end{document}